\documentclass[prb,aps,groupedaddress,amsmath,twocolumn,showpacs]{revtex4}

\usepackage{graphicx}% Include figure files
\usepackage{dcolumn}
\usepackage{bm}
\usepackage{times}
\usepackage{amsmath}
\usepackage{epsfig}
\usepackage{dcolumn}
\usepackage{amssymb}
\usepackage{color}

\begin{document}

\title{Ab-initio molecular dynamics simulation of liquid water by Quantum Monte Carlo}

\author{
Andrea Zen$^{a,e}$,
Ye Luo$^{b,c}$, 
Guglielmo Mazzola$^{b,c}$,
Leonardo Guidoni$^{a,d}$,
and
Sandro Sorella$^{b,c}$.
}
\affiliation{
$^a$ Dipartimento di Fisica, ``La Sapienza'' - Universit\`a di Roma, piazzale Aldo Moro 5, 00185, Rome, Italy; 
$^b$ SISSA -- International School for Advanced Studies, Via Bonomea 26, 34136 Trieste, Italy;
$^c$ Democritos Simulation Center CNR--IOM Istituto Officina dei Materiali, 34151 Trieste, Italy;
$^d$ Dipartimento di Scienze Fisiche e Chimiche, Universit\`a degli Studi dell' Aquila, via Vetoio, 67100, L' Aquila, Italy;
$^e$ London Centre for Nanotechnology, University College London, London WC1E 6BT, United Kingdom.
}
\email[]{
a.zen@ucl.ac.uk; 
xw111luoye@gmail.com; 
gmazzola@phys.ethz.ch; 
leonardo.guidoni@univaq.it; 
sorella@sissa.it
}

\date{\today}

\begin{abstract}
Although liquid water is ubiquitous in chemical reactions at roots of life and climate on the earth, the prediction of its properties by high-level ab initio molecular dynamics simulations still represents a  formidable task for quantum chemistry. In this article we present a room temperature simulation of liquid water based on the potential  energy surface obtained by a many-body wave function through quantum Monte Carlo (QMC) methods. The simulated properties are in good  agreement with recent neutron scattering and X-ray experiments, particularly concerning the position of the oxygen-oxygen peak in the radial distribution function, at variance of previous Density Functional Theory attempts. Given the excellent performances of QMC on large scale supercomputers, this work opens new perspectives for predictive and reliable ab-initio simulations of complex chemical  systems.
\end{abstract}

%\keywords{ Liquid Water | Radial distribution function | Hydrogen bond | Liquid simulations |  Electron correlation }

%\abbreviations{ QMC, Quantum Monte Carlo; VMC, Variational Monte Carlo; DFT, Density Functional Theory; RDF, Radial Distribution Function }

\maketitle

%\section*{Significance}
%There is no doubt about the importance of liquid water for the climate and life on Earth.
%However, correctly modeling the properties of this substance represents  still a formidable task.
%In this work we study the structural properties of liquid water, by computer simulation, with a 
%first principle approach based on a computationally demanding, but accurate and fully ab-initio, 
%stochastic solution of the Schr{\"o}dinger equation. 
%In this way it is possible to locate and determine rather accurately the peak positions 
%and shapes of the radial distribution functions characteristic of the liquid phase,
%and establish  a more direct agreement between  theory and experiments.
%\end{minipage}
%}
%\vspace{10pt}

The simulation by first principles of  
 liquid water, the key element of human life and biological processes,  has been 
a dream for several decades after the foundation of Density Functional theory (DFT), even  within
the restriction  of the Born-Oppenheimer approximation for the heavy nuclei.
Realistic 
simulations are particular important because, at  the experimental 
level, it is not possible to clarify completely 
what are the relationships between the so many different and rich 
 phases of water and  the physical interactions between 
water molecules, determined by hydrogen bonding and weak long-range van der Waals (vdW) interactions.
Moreover water is involved in many biological and chemical processes, and first principle simulations are useful to investigate and rationalize such important mechanisms.

The first attempted simulations date back to  the pioneer works by Car and Parrinello\cite{carpar,LAASONEN:1993p25911,Sprik:1996cf}, 
within an efficient ab-initio molecular dynamics (AIMD) based on DFT.
The comparison with the experiments, at that time available, provided a pretty good agreement 
with  the oxygen-oxygen (O-O) radial distribution function (RDF), as far as the positions of the peaks were concerned, 
but the overall shape given by the simulation was overstructured.
After these first studies, many other works reporting standard DFT-based simulations have been published, but  the agreement with the experimental data is still not satisfactory on many aspects.
The equilibrium density at ambient pressure (1 atm $\sim$ 10$^{-4}$ GPa), is far to be consistent with the expected one (1 gr/cm$^3$) 
though recent DFT functionals including van der Waals substantially reduce this discrepancy\cite{Ma:2012hf}.
The simulated diffusion\cite{Kuhne:2009jq} is much lower than what is expected from experiments\cite{Sit:2005cp}, and, at least in some functionals (namely, PBE and BLYP), the solidification of water occurs at a temperature which is unrealistically large ($\sim$ 410 K), so that some of  the  present DFT simulations of liquid water should be considered supercooled metastable phases\cite{Sit:2005cp,Yoo:2009bb}.

The DFT results (about which we provide a brief summary in Tab.~\ref{tab:RDF-OO-review}) appear to be substantially influenced 
by the choice of the functional\cite{VandeVondele:2005p21042,Kuhne:2009jq}, but also, within a given functional, by other details of the electronic calculations such as the pseudo-potential\cite{Sit:2005cp} and the basis set\cite{Lee:2006ge,Lee:2007hh}
-- even though all these sources of errors are perfectly controllable, including 
the size effects\cite{Grossman:2004dw,Kuhne:2009jq} and the choice of the fictitious mass in the Car-Parrinello AIMD (CPMD),\cite{Grossman:2004dw,Kuo:2004p21046} at the cost of increasing the computational cost of the simulations. --
The mostly used functionals for liquid water are those based on the generalized gradient approximation (GGA) to DFT (often PBE or BLYP density functionals),
 yielding an overstructured water at ambient conditions.
The accurate description of the exchange term by using the computationally-more-expensive hybrid functionals was  shown to improve significantly the results\cite{Todorova:2006fr,Guidon:2008jg,Zhang:2011hv,DiStasioJr:2014gy}, although they are still far from the experimental observation 
probably  due to their poor description of the long-range interaction forces. 
On the other hand, in order to overcome the well-known difficulty of DFT in describing 
long range interaction forces, 
the  inclusion of empirical dispersion terms 
has been attempted 
either by using  empirical pairwise interatomic potentials (of the $C_6 R^{-6}$ form) 
 in the total energy\cite{Grimme:2004db,Grimme:2006fc}, 
or by adopting dispersion-corrected atom-centered potentials\cite{vonLilienfeld:2004gz}. 
Another possibility is the use of the van der Waals density functionals of Dion et al.\cite{Dion:2004ce}, and further derivation based on nonlocal exchange-correlation functionals. 
All these approaches have provided remarkable  improvements in some cases\cite{Schmidt:2009p20318,Lin:2009gm,Lin:2012es,Wang:2011hx,Zhang:2011jj,DiStasioJr:2014gy},
although these methods  depend on external tunable parameters, and are strongly dependent on the functional.
Recently, Morales et al.~\cite{Morales:2014jt} have investigated the performances of several different DFT functionals versus very accurate diffusion Monte Carlo calculations, showing that the non-hybrid density functionals offers a poor description of the intramolecular potential energy surface, implying that there is still room for improvement of DFT functionals. 
% quantum nuclei
Finally, quantum effects have been shown to have an important role, as they lead to a more accurate description of the hydrogen-bond, improving the agreement with experimental data\cite{Morrone:2008kd,Habershon:2009bh,Paesani:2009ea,Paesani:2010fe,Li:2011fd,Ceriotti:2013ka}
by  broadening the  RDF.
Most of these achievements are very promising, as for instance the one reported in a recent paper by \citet{DiStasioJr:2014gy}, in which it is shown that the use of hybrid functionals, the  inclusion of vdW/dispersion interactions and the increase of the simulation temperature by 30~K, determines a remarkably improved oxygen-oxygen radial distribution function. Anyway, the issue of the choice of the DFT functional still remains controversial, because 
it is not clear if, and to which extent, a better agreement with experiments corresponds to a better 
description of the chemical and physical interactions between water molecules.

A recent accurate experiment of X-ray diffraction\cite{Skinner:2013cw}  has raised again the reliability issue of present 
%%%%XXX Skinner:2013cw doi:10.1063/1.4790861
ab-initio molecular dynamics schemes, as it was found that,  
surprisingly, the position of the first peak was shifted towards 
larger distances.
This observation is in excellent agreement with a recent extensive and independent review on the experimental structure of bulk water\cite{Soper:2013bs}.
%%%XXX Soper:2013bs http://dx.doi.org/10.1155/2013/279463
Indeed in Ref.~\citenum{Soper:2013bs} a new methodology to interpret the experimental data is employed and also shifts of the intermolecular O-O, O-H and H-H peak positions with respect to the old experimental references\cite{Soper:2000ik} are reported; see Tab.~\ref{tab:RDF-OO-review}.
%%%XXXX Soper:2000ik http://dx.doi.org/10.1016/S0301-0104(00)00179-8
These results are particularly important for ab-initio simulations because,
 the use of the PBE functional --- until recently one of the  most popular in this field --- 
is being  now replaced in favor of different functionals, like BLYP or B3LYP, 
that look clearly closer to present experiments\cite{Kuhne:2009jq}.
In other words, we believe that, in order to make some progress 
for clarifying the present discrepancies between  experiments and numerical simulations 
in this field, it is now timely to use a completely different approach for the 
following reasons:
\begin{itemize}
\item One of the main difficulties of DFT --- within its current implementation with approximate functionals --- 
is the lack of a systematic way  of improving the quality of the approximations employed, also because they cannot be validated by a variational principle as in wave-function based approaches. 
Different functionals might be more suitable to tackle different problems, and in many cases it is difficult to judge whether one functional is more accurate than another one for a given property
without knowing the experimental result. 
This means that  DFT requires alternative methods, able to validate new promising functionals, in order  to establish  properties of materials in special conditions to which experiments are not accessible, for instance at very high pressures.
\item The computing performances, especially in massively parallel architectures, are constantly 
growing with an impressive speed, as an exascale supercomputer is expected much 
before 2020, and supercomputer architectures are 
 becoming more and more suitable for statistical techniques rather 
than for deterministic methods such as DFT. On such high performance computing machines
a wave function approach based on QMC is currently becoming practical and competitive with DFT, allowing to treat geometry optimization of molecules up to 100 atoms\cite{Coccia:2014do}, vibrational properties\cite{AZenJCTC2012,Luo:2014kj} and molecular dynamics simulations\cite{Mazzola:2014dl,Mazzola:2014wc}.
\end{itemize}  

% QMC
Quantum Monte Carlo is a highly accurate wave-function-based approach for electronic structure calculations\cite{Foulkes:2001p19717}, that has been also recently extended for ab-initio simulations\cite{Grossman:2005p21519,attaccalite,mazzola_finite-temperature_2012,reboredo2014generalizing,ceperley_penalty_1999,Mazzola:2014dl,Mazzola:2014wc}.
% in this work
In this work we have employed the first ab-initio molecular dynamics simulation of liquid water based entirely  on QMC.
We adopt the Born-Oppenheimer approximation, neglecting the quantum effects 
on ions, and apply the  variational Monte Carlo (VMC) approach using as an ansatz  
 a Jastrow Slater many-body wave function.
Even though we have used the simplest QMC approach, a significant 
improvement in the description of liquid water has been achieved.
In particular we have obtained that the O-O RDF, $g_{\rm OO}(r)$, is considerably less structured compared with  DFT calculations of the same type (with no proton quantum effects).
Moreover, it is also worth to emphasize that the position of the first peak is now
in perfect agreement with the most recent and accurate 
experiments, a fact that was indeed found with a simulation dated 
before the new experimental data were distributed\cite{qmcwastrue}.  %%GGG
%\texttt{http://www.int.washington.edu/talks/WorkShops/int_13_2a/People/Sorella_S/Sorella.pdf}
%\texttt{http://www.int.washington.edu/talks/WorkShops/int\char`_13\char`_2a/People/Sorella\char`_S/Sorella.pdf}

%XXX OUTLINE XXX
The paper is organized as follows: 
in Section \ref{sec.methods} we  describe the methodological aspects of the QMC-based molecular dynamics simulation, and we provide the details about the variational ansatz used for the dynamics and its expected accuracy; in Section \ref{sec.results} we discuss the results obtained simulating the liquid water by the QMC-based MD, and in Section \ref{sec.discussion} we discuss these results and draw the concluding remarks.

%%% COMPARISON OTHER METHODS
\begin{table*}
\caption{
Structural properties (position and height of the nearest neighbor maximum in $g_{OO}(r)$ and minimum) and computational details for several {\em ab-initio} simulations of liquid water in ambient conditions, as reported in recent literature for DFT-based molecular dynamics with PBE or BLYP functionals, in comparison with experiments and VMC-based results obtained in this work. 
The dynamics column indicates if results are obtained using:
Born-Oppenheimer (BO) molecular dynamics,
Car-Parrinello (CP) molecular dynamics,
Second generation Car-Parrinello (2GCP) molecular dynamics\cite{Kuhne:2007df},
Monte Carlo (MC) sampling,
or the Langevin dynamics (LD) adopted in this work.
The number in the parenthesis in the CP dynamics identifies the value of the fictitious mass of the electron $\mu$.
The sampling time correspond to the production run.
\\
}\label{tab:RDF-OO-review}
%{\tiny
%{\scriptsize
%{\footnotesize
%{\small
%%% START TABEXEL
\begin{tabular}{l c  c c c c c c c c c c c c c }																							
\hline																							
Ref.	&	N	&	System	&	Method	&	Dynamics	&	Ensemble	&	Sampling	&	$T_{ion}$ [K]	&	$r_{MAX}$	&	$g_{MAX}$	&	$r_{min}$	&	$g_{min}$	\\
\hline																							
\citet{Grossman:2004dw}	&	32	&	H$_2$O	&	BLYP	&	CP($340$)	&	NVE	&	20 ps	&	285.9	&	2.73	&	3.65	&	3.32	&	0.40	\\
\citet{Grossman:2004dw}	&	32	&	D$_2$O	&	BLYP	&	CP($340$)	&	NVE	&	20 ps	&	297.5	&	2.73	&	3.60	&	3.33	&	0.39	\\
\citet{Kuo:2004p21046}  \footnote{ Simulation: CPMD-NVE-400. }	&	64	&	H$_2$O	&	BLYP	&	CP($400$)	&	NVE	&	20 ps	&	314	&	2.76	&	2.90	&		&		\\
\citet{Kuo:2004p21046}  \footnote{ Simulation: CPMD-NVE-BO. }	&	64	&	H$_2$O	&	BLYP	&	BO	&	NVE	&	10 ps	&	323	&	2.76	&	3.00	&		&		\\
\citet{Kuo:2004p21046}  \footnote{ Average of CP2K-MC-NVT-1 and CP2K-MC-NVT-2. }	&	64	&	H$_2$O	&	BLYP	&	MC	&	NVT	&		&	300	&	2.76	&	2.95	&		&		\\
\citet{Lee:2006ge}	&	32	&	D$_2$O	&	BLYP	&	CP($500$)	&	NVT	&	30 ps	&	300	&	2.77	&	2.90	&		&		\\
\citet{Kuhne:2009jq}	&	64	&	H$_2$O	&	BLYP	&	2GCP	&	NVT	&	30 ps	&	300	&	2.79	&	2.92	&	3.33	&	0.57	\\
\citet{Lin:2012es}	&	64	&	D$_2$O	&	BLYP	&	CP($600$)	&	NVE	&	40 ps	&	319	&	2.77	&	2.86	&	3.31	&	0.66	\\
\hline																							
\citet{Grossman:2004dw}	&	32	&	H$_2$O	&	PBE	&	CP($340$)	&	NVE	&	20 ps	&	290.8	&	2.71	&	3.46	&	3.30	&	0.41	\\
\citet{Grossman:2004dw}	&	54	&	H$_2$O	&	PBE	&	CP($340$)	&	NVE	&	12 ps	&	298	&	2.73	&	3.75	&	3.36	&	0.78	\\
\citet{Schwegler:2004ii}	&	54	&	H$_2$O	&	PBE	&	CP($340$)	&	NVE	&	19.8 ps	&	296	&	2.69	&	3.65	&	3.32	&	0.37	\\
\citet{Schwegler:2004ii}	&	54	&	H$_2$O	&	PBE	&	BO	&	NVE	&	20.5 ps	&	306	&	2.72	&	3.83	&	3.25	&	0.33	\\
\citet{Kuhne:2009jq}	&	64	&	H$_2$O	&	PBE	&	2GCP	&	NVT	&	250 ps	&	300	&	2.73	&	3.25	&	3.28	&	0.44	\\
\citet{Lin:2012es}	&	64	&	D$_2$O	&	PBE	&	CP($600$)	&	NVE	&	40 ps	&	314	&	2.72	&	3.19	&	3.27	&	0.43	\\
\citet{DiStasioJr:2014gy}	&	64	&	D$_2$O	&	PBE	&	CP($300$)	&	NVT	&	$>$20 ps	&	300	&	2.69	&	3.28	&	3.28	&	0.37	\\
\hline																							
\citet{DiStasioJr:2014gy}	&	64	&	D$_2$O	&	PBE0	&	CP($300$)	&	NVT	&	$>$20 ps	&	300	&	2.71	&	2.96	&	3.30	&	0.53	\\
\citet{DiStasioJr:2014gy}	&	64	&	D$_2$O	&	PBE+vdW	&	CP($300$)	&	NVT	&	$>$20 ps	&	300	&	2.71	&	2.99	&	3.27	&	0.54	\\
\citet{DiStasioJr:2014gy}	&	64	&	D$_2$O	&	PBE0+vdW	&	CP($300$)	&	NVT	&	$>$20 ps	&	300	&	2.72	&	2.76	&	3.31	&	0.70	\\
\citet{DiStasioJr:2014gy}	&	128	&	D$_2$O	&	PBE0+vdW	&	CP($300$)	&	NVT	&	$>$20 ps	&	330	&	2.74	&	2.51	&	3.33	&	0.84	\\
\hline																							
this work	&	32	&		&	VMC	&	LD	&	NVT	&		&	300	&	2.80	&	3.36	&	3.32	&	0.69	\\
\hline																							
\multicolumn{7}{l}{Experiment: \citet{Soper:2013bs} (2013) }		&	298	&	2.79	&	2.49	&		&		\\
\multicolumn{7}{l}{Experiment: \citet{Skinner:2013cw} (2013) }	&	298	&	2.80(1)	&	2.57(5)	&	3.45(4)	&	0.84(2)	\\
\hline																							
\end{tabular}																							
%%% END TABEXEL
%}
\end{table*}

\section{Methodological aspects of the QMC-based Molecular Dynamics} \label{sec.methods}

The molecular dynamics, driven by quantum Monte Carlo forces, was
introduced recently for the simulation of liquid hydrogen at high pressures\cite{Mazzola:2014dl,Mazzola:2014wc} and to obtain vibrational properties of molecular systems\cite{Luo:2014kj}.
At fixed ion coordinates ${\bf R}$ the many-body wave function depends on the 
$N-$ electronic positions $x=\{ \vec r_1, \vec r_2 , \cdots \vec r_N \}$ by means of 
the following Jastrow Slater (JSD) ansatz: 
\begin{equation}  \label{thewf}
\Psi_\textrm{JSD} = J * \Psi_\textrm{SD} \,,
%\Psi_\textrm{JSD} =  e^{U} * \Psi_\textrm{SD} \,
\end{equation}
where  $\Psi_\textrm{SD}$ is a single Slater determinant and  $J= e^{U}$ is the Jastrow factor.
The Jastrow factor is a  symmetric positive function of the electronic positions that depends on the inter-particle  distances, and   describes the dynamical correlation among electrons. It is also particularly useful because, already in its simplest form, makes it possible to fulfill the electron-electron and electron-nucleus cusp conditions\cite{Foulkes:2001p19717,Drummond:2004p18505,zenwater}. 
So far the JSD ansatz can be efficiently simulated  within a quantum Monte Carlo\cite{Foulkes:2001p19717,zenwater} approach, that introduces no other bias than the statistical error, systematically vanishing with the simulation length.

An extensive discussion about the Langevin dynamics that we have used to integrate the equations of motions, when dealing with error affected nuclear force evaluations (because coming from QMC methods), has been already provided in Ref.~\citenum{Luo:2014kj}, and the interested reader can refer to that. 
In the following we instead want to discuss in much more details the variational ansatz that we have used for the specific system here under consideration. 
The choice of the ansatz (such as the functional form of the Jastrow and the basis set used) is of major importance in order to have accurate results within QMC calculations, as well as for any other wave-function-based method. 
Thus, in subsections \ref{sec:det-wf} and \ref{sec:jas-wf}  we provide all the details concerning respectively the determinantal and Jastrow part of the wave function used in the VMC-based molecular dynamics.
This wave function is the result of an extensive work for having a wave function that is accurate enough to provide reliable VMC results, and that is sufficiently compact (namely, that has a reasonably small number of parameters) that it can be stably and efficiently optimized in any MD step.

In order to show the quality of our approach, we have considered the water dimer, the simplest system in which the hydrogen bond is present, for which we report and compare, in subsection \ref{sec:dimer-test}, a number of tests of different basis sets and ansatzes, both at the variational and the fixed-node lattice regularized diffusion Monte Carlo scheme\cite{Casula:2005p14138,Casula:2010p14082} (LRDMC).
The latter method projects our approximate  wave function to the exact ground state, with the approximation that the nodes of the wave function are pinned to the initial value determined by our ansatz. 
Among several advantages, LRDMC guarantees the full variational upper bound property of the energy, even when pseudo-potentials are used\cite{Casula:2005p14138,Casula:2010p14082}.

Finally, the motivations for our final choice of the ansatz and basis set are reported in subsection \ref{sec.thechoice}.

\subsection{ Determinant part $\Psi_\textrm{SD}$ and its basis set }\label{sec:det-wf}

In the Slater determinant $\Psi_\textrm{SD}$, the double occupied molecular orbitals $\Psi_i$, with index $i=1,\ldots,N/2$ ($N$ is the number of electrons),
are a linear combination of the localized atomic hybrid orbitals\cite{zenwater}:
\begin{equation}
\Psi_i({\bf r}) = \sum_{a=1}^{M} \sum_{\mu_a=1}^{L_a} c_i^{a,\mu_a} \Phi_{a,\mu_a}({\bf r})
\end{equation}
where $\Phi_{a,\mu_a}$ is the $\mu_a$-th atomic hybrid orbital of atom $a$, centered around the position ${\bf R}_a$ of nucleus $a$, $M$ is the total number of atoms, 
$L_a$ is the number of atomic hybrid orbitals used for atom $a$, for a total of $L=\sum_a L_a$ hybrid orbitals in the overall system, 
and the $M\times L$ coefficients $c_i^{a,\mu_a}$ are variationally optimized.
% coefficient optimization
The optimization is performed by using the correspondence between the single Slater determinant written in terms of molecular orbitals, and a truncated antisymmetrized geminal power (AGPn)\cite{Marchi:2009p12614,zenwater} with $n=N/2$, with a geminal
\begin{equation}
g({\bf r}_1,{\bf r}_2) = 
  \sum_{a,b}^{M} \sum_{\mu_a}^{L_a} \sum_{\mu_b}^{L_b}  
  g^{a,b}_{\mu_a,\mu_b}
  \Phi_{a,\mu_a}({\bf r}_1) \Phi_{b,\mu_b}({\bf r}_2) \,
\end{equation}
those $L\times L$ parameters $g^{a,b}_{\mu_a,\mu_b}$ are related to $c_i^{a,\mu_a}$ by the relation:
\begin{equation}
g^{a,b}_{\mu_a,\mu_b} = \sum_{i=1}^{N/2} c_i^{a,\mu_a} c_i^{b,\mu_b} \,.
\end{equation}
The present formulation is adopted in the {\em TurboRVB} code,\cite{TurboRVB} because in this 
way it is much simpler to satisfy symmetry properties (e.g. a spin singlet implies that the 
matrix $g$ is symmetric) and to decrease  the number of parameters by disregarding, during the 
optimization, those  matrix elements,   corresponding to atomic centers 
located at large distance each other (see later).
Therefore, the parameters actually  optimized in this approach are the $g^{a,b}_{\mu_a,\mu_b}$. They are then used to obtain the molecular orbital coefficients $c_i^{a,\mu_a}$ via the diagonalization described in Refs.~\citenum{Marchi:2009p12614,zenwater}.
This choice gives a very important technical advantage for systems of large sizes as the one considered in this work.
In particular, in order to decrease the total number of variational parameters, 
 we have fixed to zero all the coefficients  $g^{a,b}_{\mu_a,\mu_b}$ connecting the atoms $a$ and $b$ that are at a distance 
$R_{ab} = \| {\bf R}_a - {\bf R}_b \| $ 
larger than an appropriately chosen cut-off $R_\textrm{MAX}$.
We will see in the following sections that this choice does not significantly affect the accuracy of the calculation, 
but it is important because a much smaller number of variational parameters guarantees 
a stable and efficient wave function optimization.

% BASIS SET DETERMINANT
The $\mu_a$-th atomic hybrid orbital $\Phi_{a,\mu_a}$ of the atom $a$ is expressed as a linear combination of all the uncontracted orbitals 
$\phi_{a,l,k}$
introduced to describe the system:
\begin{equation}
\Phi_{a,\mu_a}(\textbf{r}) = 
  \sum_{l=0}^{l_{M}(a)} 
    \sum_{m=-l}^{+l} 
    \sum_{k=1}^{K_{M}(a,l)} 
        h_{a,l,m,k} \, \phi_{a,l,k}(r) \, Z_{l,m}(\Omega) \,,
\end{equation}
where $l$ is the azimuthal quantum number, running from zero to the maximum angular momentum $l_{M}(a)$ of atom $a$;
$m$ is the magnetic quantum number;
$k$ labels the $K_M(a,l)$ orbitals of angular momentum $l$ of the atom $a$ in the chosen basis set;
$Z_{l,m}(\Omega)$ is the real spherical harmonic and $r = \| {\bf r} \| $ is the distance of the electron from the nucleus $a$.
The coefficients $h_{a,l,m,k}$ are  parameters that are variationally optimized.

% uncontracted orbitals
The uncontracted orbitals implemented in {\em TurboRVB} code\cite{TurboRVB} are essentially Gaussian (GTO) or Slater type orbitals (STO), with some possible modifications (as in the case of the STO s-orbital, modified in a way to avoid the electron-nucleus cusp, 
already satisfied by the chosen Jastrow factor) or 
generalizations (as for the case of the $r^2*\textrm{GTO}$ s-orbital), allowing an 
improved description of  the orbital shape with the minimum possible number of variational parameters.
In this work the orbital functions that we have used are the following (for open systems):
\begin{itemize}
% s
\item {\bf s}-orbitals ({\em i.e.} $l=0$): 
\begin{eqnarray}
\phi_{\bf s}^\textrm{STO}(r) & \propto & (1+\zeta r) e^{-\zeta r} \\
\phi_{\bf s}^\textrm{GTO}(r) &\propto & e^{-\zeta r^2} \\
\phi_{\bf s}^{r^2*\textrm{GTO}}(r) &\propto & r^{2} e^{-\zeta r^2}
\end{eqnarray}
% p
\item {\bf p}-orbitals ({\em i.e.} $l=1$): 
\begin{eqnarray}
\phi_{\bf p}^\textrm{STO}(r) & \propto & r e^{-\zeta r} \\
\phi_{\bf p}^\textrm{GTO}(r) &\propto & r e^{-\zeta r^2} 
\end{eqnarray}
% d
\item {\bf d}-orbitals ({\em i.e.} $l=2$): 
\begin{eqnarray}
\phi_{\bf d}^\textrm{STO}(r) & \propto & r^2 e^{-\zeta r} \\
\phi_{\bf d}^\textrm{GTO}(r) &\propto & r^2 e^{-\zeta r^2} 
\end{eqnarray}
\end{itemize}
whereas for systems with periodic boundary conditions (PBC), 
as described in Refs.~\citenum{AttaccalitePhD,Sorella:2011p24127} and already used in Refs.~\citenum{attaccalite,Mazzola:2014dl,Sorella:2011p24127},
the orbital functions are slightly modified, namely the Cartesian distance $\mathbf r$ is replaced by a simple periodic function $
\tilde{ \mathbf r}(\mathbf r)$ that
takes into account the appropriate periodicity of the box. By consequence, 
 also the distances $r=\| \mathbf r \|$ are replaced by new distances $\tilde r = \| \tilde{ \mathbf r} \|$ and the normalization coefficients are correspondingly changed.  
In particular, we have used here the following substitution rule to modify an orbital used 
for  open systems to PBC with box of of lengths $(L_x, L_y,L_z)$:
\begin{equation}\label{eq:open2pbc}
(x,y,z) \to (\tilde x, \tilde y, \tilde z) = \left(
{L_x\over \pi}\sin{\pi x \over L_x} , 
{L_y\over \pi}\sin{\pi y \over L_y} , 
{L_z\over \pi}\sin{\pi z \over L_z} \right)
\, .
\end{equation}
Other parametric forms for the atomic orbitals exist, see for instance \citet{Petruzielo:2011p24345}, but are not used in this work.
Each of the uncontracted orbitals described above depends parametrically only on the value of the $\zeta$ in the exponent, that can be optimized as all the other variational parameters within our VMC calculations, see Refs.~\citenum{Casula:2003p12694,zenwater}.
% Z optimized for the dimer
In order to enhance the stability of the wave function optimization during the dynamics, and to reduce the computational effort, we have optimized the values of the $\zeta$ exponents for the water dimer, namely the smallest system with an hydrogen bond, and we have kept these exponents fixed in the VMC-based molecular dynamics.

The atomic basis set used in this work, including the orbital types and the exponent values for the oxygen and the hydrogen atoms, are specified in Tab.~\ref{tab:wfpar}.

\subsection{ Jastrow factor and its basis set }\label{sec:jas-wf}
% BASIS SET JASTROW

In the VMC-based molecular dynamics we used the Jastrow factor 
$$ J = \exp({U_{en} + U_{ee} + U_{een}}) , $$
that involves:
the one-electron interaction term $U_{en}$, 
the homogeneous two electron interaction term $U_{ee}$,
and the inhomogeneous two-electron interaction term $U_{een}$, representing an electron-electron-nucleus function.
They are defined as follows:
\begin{eqnarray}
\label{eq.Jen}
U_{en}( \bar{\textbf{r}}) &=& 
\sum_{i}^N  \sum_{a}^M  \left[
  - %(2 Z_a)^{3\over4} u_1\left( \sqrt[4]{2 Z_a} r_{ia} \right) 
    Z_a { 1-e^{- b_1 \sqrt[4]{2 Z_a} r_{ia}} \over b_1 \sqrt[4]{2 Z_a} } 
    \right. \nonumber \\
  &&  
  + \left. \sum_{\mu_a}
        f^a_{\mu_a} \chi_{a,\mu_a}(\textbf{r}_{i}) 
      \right]
\\
\label{eq.Jee}
U_{ee}\left(\bar{\textbf{r}}\right) &=& 
  \sum_{i<j}^N \left[
%    { 1-e^{- b_2 r_{ij}} \over 2 b_2 }       % short range 2-body
    { r_{ij}  \over { 2 ( 1 + b_2 r_{ij} ) } }    % long range 2-body
    \right]
\\
\label{eq.Jeen}
U_{een}\left( \bar{\textbf{r}} \right) &=& 
\sum_{i<j}^N   \sum_{a}^M 
      \sum_{\mu_a,\nu_a} \left[
        \bar f^{a}_{\mu_a,\nu_a} 
        \right.
   \nonumber \\ && \left.
          \chi_{a,\mu_a}\left(\textbf{r}_{i}\right) \;
          \chi_{a,\nu_a}\left(\textbf{r}_{j}\right)
  \right] ,
\end{eqnarray}
where 
the vector 
$\textbf{r}_{ia}=\textbf{r}_i-\textbf{R}_a$
is the difference between the position of the nucleus $a$ and the electron $i$,  
$r_{ia}$ is the corresponding distance,
$r_{ij}$ is the distance between electrons $i$ and $j$,  
%$ u_1(x)= { 1-e^{- b_1 x} \over 2 b_1 } $,
$Z_a$ is the electronic charge of the nucleus (or pseudo nucleus) $a$,
$\chi_{a,\mu_a}$ are the atomic orbitals of nucleus $a$, %  (they are similar to the ones used for the AGP), 
and $b_1$, $b_2$, 
$f^a_{\mu_a}$, 
$\bar f^a_{\mu_a,\nu_a}$, 
%$\tilde f^{a,b}_{\mu_a,\mu_b}$ 
are variational parameters.
The leading contribution for the description of electronic correlation is given by  $U_{ee}$, but also the inhomogeneous two-electron interaction term $U_{een}$ is important, because they can improve the charge distribution\cite{Sorella:2007p12646}.

At variance of a previous QMC study\cite{zenwater},  in the present VMC-based molecular dynamics, we did not include in the Jastrow factor the electron-electron-nucleus-nucleus term  
\begin{equation}
\label{eq.Jeenn}
U_{eenn}\left( \bar{\textbf{r}}\right) = 
\sum_{i<j}^N  \sum_{a \neq b}^M  \left[
      \sum_{\mu_a} \sum_{\mu_b}
        \tilde f^{a,b}_{\mu_a,\mu_b} 
          \chi_{a,\mu_a}\left(\textbf{r}_{i}\right)
          \chi_{b,\mu_b}\left(\textbf{r}_{j}\right)
  \right] ,
\end{equation}
that could further improve the description of the long range electron correlation and the charge distribution, but requires a number of coefficients $f^{a,b}_{\mu_a,\mu_b}$ that grows quadratically with the number of atomic orbitals. Thus this term  is not computationally affordable for a system as large as the ones  considered here.
On the other hand, the functional form of the homogeneous two-electron interaction term $U_{ee}$ that we are using here has a long-range correlation, that satisfactorily recovers a  part of the 
 correlation implied  by the $U_{eenn}$ term, and reproduces the correct bulk properties.

% pbc
In presence of PBC, the coordinates and the distances are modified in order to fulfill the periodicity of the system, as discussed in the previous section. In the Jastrow factor, not only the localized atomic orbitals are modified, but also the homogeneous term in the electron-nucleus and electron-electron terms are obviously affected by this change.

% values
The values of $b_1$ and $b_2$ parameters are optimized during the dynamics, and their optimized values  are around $\simeq 1$ and $\simeq 0.5$, respectively. 
We have considered uncontracted atomic orbitals $\chi_{a,\mu_a}$ of the GTO  type for the inhomogeneous terms, and the values of the exponents have been optimized for the water dimer and kept fixed during the VMC-based dynamics, as for the determinant case.  
The atomic basis set used in this work, including the orbital types and the exponent values for the oxygen and the hydrogen atoms, are specified in Tab.~\ref{tab:wfpar}

\begin{table}
\caption{ Basis set parameters in the wave-function used for the VMC-based molecular dynamics. The two core electrons of the oxygen atoms have been described using  the scalar-relativistic energy consistent pseudopotential of Burkatzki {\it et al.}\cite{Burkatzki:2007p25447} %Filippi
}
\label{tab:wfpar}
%{\tiny
%{\scriptsize
%{\footnotesize
%{\small
\begin{tabular}{ | l c | l c |  }  
\multicolumn{4}{c}{} \\				
\multicolumn{4}{c}{ {\bf Determinant part}  }			\\
\multicolumn{4}{c}{ $R_{MAX}=4.5$ }			\\

\multicolumn{2}{c}{\bf \em O: (5s,5p,1d)/\{5\} }			&	
\multicolumn{2}{c}{\bf \em H: (3s,1p)/\{3\} }			\\

\hline
type	&	$\zeta$	&	type	&	$\zeta$	\\
\hline														
s - STO	&	2.037	&	s - STO	&	1.572	\\
s - $r^2 *$GTO	&	1.128	&	s - GTO	&	0.086	\\
s - GTO	&	0.214	&	s - GTO	&	2.176	\\
s - GTO	&	0.736	&		&		\\
s - GTO	&	3.617	&		&		\\
p - STO	&	1.199	&	p - STO	&	1.112	\\
p - GTO	&	0.433	&		&		\\
p - GTO	&	1.408	&		&		\\
p - GTO	&	4.183	&		&		\\
p - GTO	&	10.380	&		&		\\
d - STO	&	1.202	&		&		\\
\hline			
\multicolumn{4}{c}{} \\				
\multicolumn{4}{c}{ {\bf Jastrow part}  }	 \\						

\multicolumn{2}{c}{\bf \em O: (3s,2p) }			&	
\multicolumn{2}{c}{\bf \em H: (2s,2p) }			\\

\hline
type	&	$\zeta$	&	type	&	$\zeta$	\\
\hline	
s - G	&	2.022	&	s - G	&	1.648	\\
s - G	&	0.507	&	s - G	&	0.051	\\
s - G	&	0.231	&		&		\\
p - G	&	0.747	&	p - G	&	0.075	\\
p - G	&	0.084	&	p - G	&	0.697	\\
\hline							
\end{tabular}
%}
\end{table}

% dimer dissociation
\begin{figure}
\begin{center}
{\large A}
\includegraphics[width=.35\textwidth]{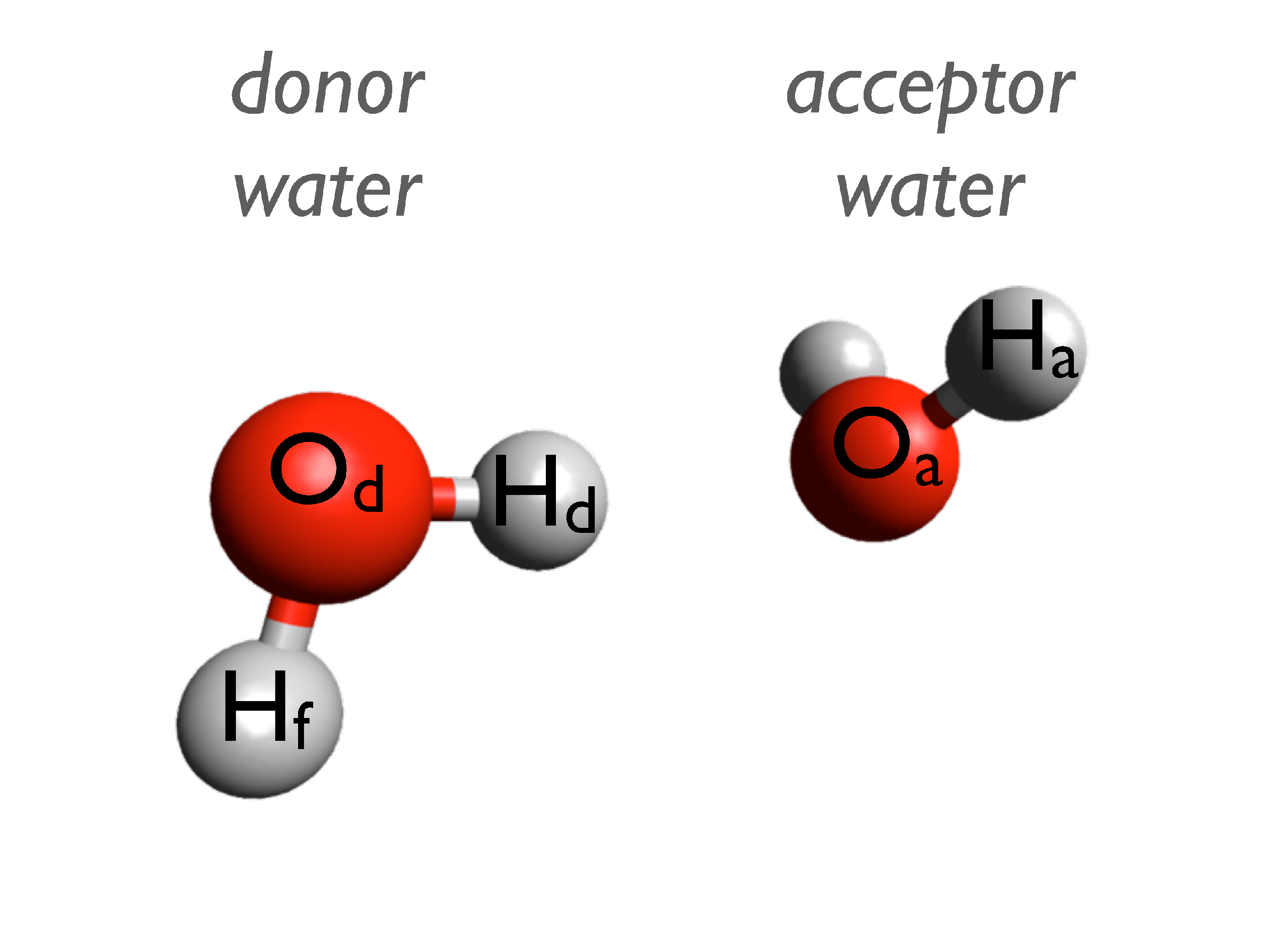} \\
{\large B}
\includegraphics[width=0.9\columnwidth]{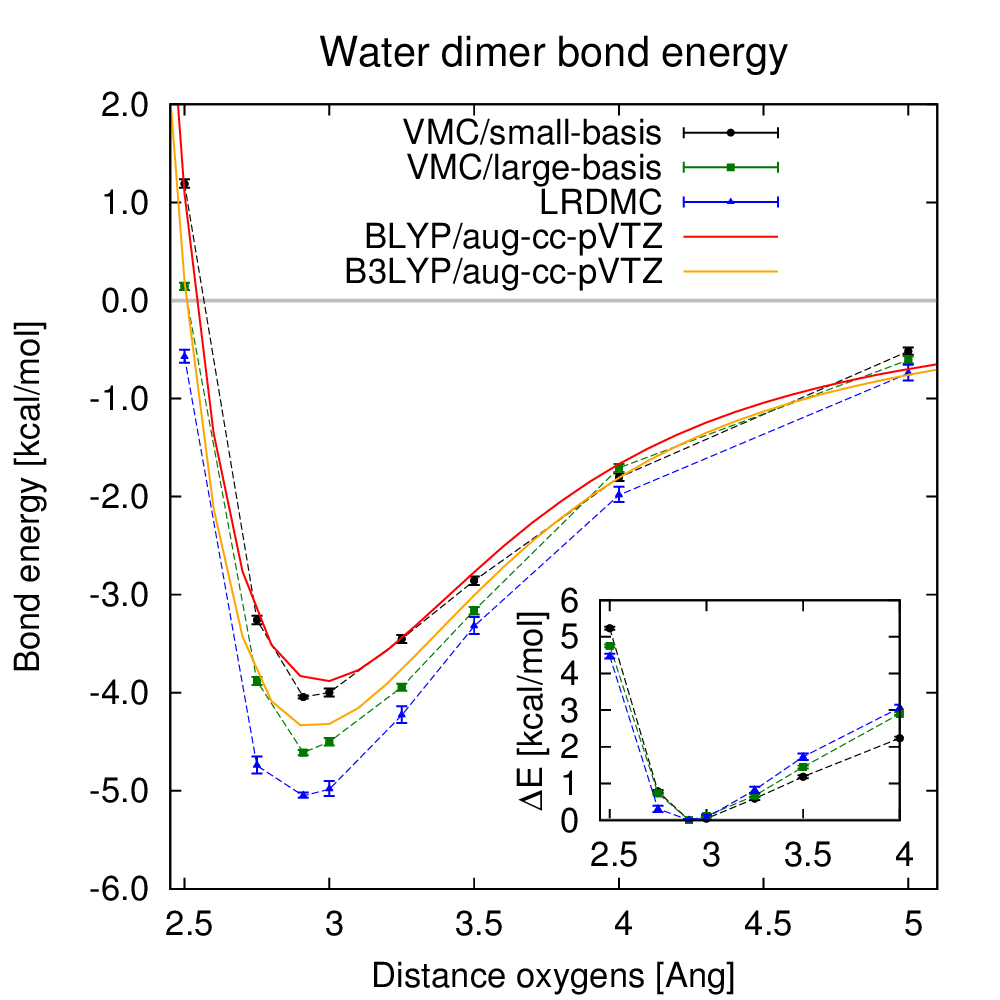}
\caption{
Panel A: 
Bonding geometry of the water dimer in its structural minimum. 
Panel B:
Dissociation energy of the water dimer, plotted as a function of the oxygen-oxygen distance, studied with  VMC (the small-basis is the one used for the dynamics), and LRDMC (that is almost independent on the choice of the small- or large-basis, see Tab.~\ref{tab:dimer-bonding} and Fig.~\ref{fig:lrdmc}). For a comparison, we report also the dissociation curve for DFT/BLYP and DFT/B3LYP, both with aug-cc-pVTZ basis set. 
Inset: energy difference with the water dimer in its equilibrium configuration.
Further details are reported in the text.
}
\label{fig:W2dissoc}
\end{center}
\end{figure}

%%%%%%%%%%%%%%%%%%%
%%% WATER DIMER %%%
%%%%%%%%%%%%%%%%%%%

\subsection{ The water dimer as a test case for the wave function ansatz and basis set 
used in this work }\label{sec:dimer-test}

In order to test the reliability of our VMC approach, and in particular of the wave function ansatz described in the previous section and used for the VMC-based molecular dynamics, we have performed several tests on the water dimer. 
In the water dimer, the structural minimum, reported in Fig.~\ref{fig:W2dissoc}A, corresponds to a configuration where one water is in the  {\em donor} configuration, and one of its hydrogens (the donor hydrogen $H_d$) is shared with the other water molecule, the {\em acceptor} water, forming the {\em hydrogen bond}.
An accurate description of the hydrogen bond is the main ingredient for the description also of the liquid water and of the ice, thus the water dimer represents a simple meaningful system to check the accuracy of our approach.

%%% TAB. BASIS SET
\begin{table*}[btp]
\caption{
Basis set for water dimer tests.
The number of atomic hybrid orbitals are reported in brace parenthesis,
whereas ``unc.'' stands for uncontracted orbitals.
In the homogeneous electron-nucleus or electron-electron terms of the Jastrow factor, $U_{en}$ and $U_{ee}$, we indicate with
``short'' [range] the functional form 
${{ 1-\exp(-br)} \over 2b}$,   % short: -6, and 1-body of -15
and with ``long'' [range] the functional form  
$r\over 2(1+br)$.   % long: -5, and 2-body of -15
We use the symbol ``=''
to indicate that the entry is the same of the previous line.
The basis sets are ordered in increasing order of number of variational parameters, thus of expected accuracy (with the exception of ``large-basis4'' and ``large-basis5'', that are equivalent in size).
The parameters $R_\textrm{MAX}$ indicates the cut-off distance (expressed in atomic units) for the coefficients $g^{a,b}_{\mu_a,\mu_b}$ relative to atoms $a$ and $b$ appearing in the determinantal part of the wave function, as described in Section~\ref{sec:det-wf}; in case the distance $R_{ab} = \| {\bf R}_a - {\bf R}_b \| > R_\textrm{MAX}$, $g^{a,b}_{\mu_a,\mu_b}$ is set to zero and not optimized.
Similarly, $R_\textrm{J-MAX}$ is the distance cut-off for the $\tilde f^{a,b}_{\mu_a,\mu_b}$ coefficients in the electron-electron-nucleus-nucleus term $U_{eenn}$ in the Jastrow factor, see Eq.~\ref{eq.Jeenn}; thus a value of $R_\textrm{J-MAX}=0$ corresponds to no $U_{eenn}$ term in the Jastrow factor.
In the $U_{een}$ and $U_{eenn}$ columns we indicate the kind of contraction of the orbital functions used for the oxygen and hydrogen atoms.
We implicitly assume that these basis sets are used with the 
scalar-relativistic energy consistent pseudopotential of Burkatzki {\it et al.}\cite{Burkatzki:2007p25447} for the two core electrons of the oxygen atom. 
\\
}
\label{tab:basis-dimer}
%%%
\begin{tabular}{ | l |  c c c | c c  c c c   c c | }																					
\hline																					
NAME	&	\multicolumn{3}{c|}{\bf Determinant}					&	\multicolumn{7}{c|}{\bf Jastrow }													\\
	&	oxygen	&	hydrogen	&	$R_\textrm{MAX}$	&	oxygen	&	hydrogen	&	$U_{een}$	&	$U_{eenn}$	&	$R_\textrm{J-MAX}$	&	$U_{en}$	&	$U_{ee}$	\\
\hline																					
small-basis1	&	(5s,5p,1d)/\{5\}	&	(3s,1p)/\{3\}	&	4.5	&	(3s,2p)	&	(2s,2p)	&	unc.	&	no	&	0	&	short	&	long	\\
small-basis2	&	=	&	=	&	=	&	=	&	=	&	=	&	O:\{1\} H:\{1\}	&	$\infty$	&	=	&	=	\\
small-basis3	&	=	&	=	&	=	&	(3s,2p,1d)	&	=	&	=	&	no	&	0	&	=	&	=	\\
small-basis4	&	=	&	=	&	=	&	=	&	=	&	=	&	O:\{1\} H:\{1\}	&	$\infty$	&	=	&	=	\\
small-basis5	&	=	&	=	&	=	&	=	&	=	&	=	&	unc.	&	$\infty$	&	=	&	=	\\
\hline																					
large-basis1	&	(9s,9p,3d,2f)/\{12\}	&	(7s,6p,2d)/\{4\}	&	$\infty$	&	(5s,4p,2d,1f)	&	(3s,2p,1d)	&	unc.	&	no	&	$\infty$	&	short	&	long	\\
large-basis2	&	=	&	=	&	=	&	=	&	=	&	=	&	O:\{1\} H:\{1\}	&	=	&	=	&	=	\\
large-basis3	&	=	&	=	&	=	&	=	&	=	&	=	&	O:\{8\} H:\{4\}	&	=	&	=	&	=	\\
large-basis4	&	=	&	=	&	=	&	=	&	=	&	=	&	unc.	&	=	&	=	&	=	\\
large-basis5	&	=	&	=	&	=	&	=	&	=	&	=	&	unc.	&	=	&	=	&	short	\\
\hline																					
\end{tabular}																					
%%%
\end{table*}

% tested basis
The main ingredient for an {\em ab-initio} treatment of the hydrogen bond is the possibility 
 to describe the {\em dynamical electronic correlation} in the system.
This is mainly contained, within our VMC-based approach,  in the Jastrow term, whereas the determinantal part of the wave function is more important in the description of strong covalent bonds\cite{Marchi:2009p12614,Braida:2011p27951,Anderson:2010p25299,Zimmerman:2009hh} %, such as the ones present in an isolated water molecule, 
and transition states\cite{Zen:2014dh,Zen:2015eu}.
Therefore, we have considered Jastrow terms of increasing size and complexity,\footnote{We have to remember that in a wave function with a large number of parameters, the variational optimization becomes more challenging, and sometimes unstable.} and only two different basis sets for the determinantal part of the wave function, that have been indicated by ``small-basisX'' and ``large-basisX'', where ``X'' is a number referring to the size of the Jastrow.
The considered basis sets are defined in Tab.~\ref{tab:basis-dimer}.
In all the reported calculations, the  
scalar-relativistic energy consistent pseudopotential (ECP) of Burkatzki {\it et al.}\cite{Burkatzki:2007p25447} %Filippi
has been adopted in order to describe the two core electrons of the oxygen atoms.
We have already shown in Ref.~\citenum{zenwater} that the use of ECP does not significantly 
affect the accuracy of  the water monomer, as compared with a corresponding 
all electrons calculation. Moreover,  ECP  is also particularly convenient as 
compared with other choices of pseudo-potentials, because it is very favorable from a computational point of view. 
Notice that the basis that was used for the VMC-based molecular dynamics is the ``small-basis1''.
We also observe that all the basis defined in  Tab.~\ref{tab:basis-dimer} and here tested have an uncontracted electron-electron-nucleus term $U_{een}$. This is due to the fact that we have observed that the contraction of the basis, in the  $U_{een}$ term, is typically not convenient, because the number of parameters in $U_{een}$ grows only linearly with the size of the basis and of the system, see Eq.~\ref{eq.Jeen}, and the computational gain with an uncontracted basis is typically important.
On the other hand, for the $U_{eenn}$ term the number of parameters grows quadratically with the size of the basis set and of the system, see Eq.~\ref{eq.Jeenn}, thus the use of an uncontracted basis turns to be  computationally very expensive, and unfeasible for very large systems. 
We have experienced  that the use of hybrid orbitals\cite{zenwater} to contract the orbitals in the $U_{eenn}$ is a very promising strategy, because it allows to minimize the number of parameters  without affecting too much the variational flexibility of our Jastrow factor. 
However, we have chosen for the dynamics,  a basis without the $U_{eenn}$ term, because  the possible improvement in  accuracy can be obtained only with further 
$\simeq 5000$ parameters for a system as large as 32 water molecules. Therefore 
 in  this work we have finally chosen the simplest basis, 
that guarantees  a very stable and efficient optimization during the dynamics.

\begin{table*}
\caption{ 
Energy, variance and dipole of the water monomer and dimer (respectively in the experimental and CCSD(T)/cc-pVQZ optimized\cite{Jurecka:2006bg,S22-database} nuclear configuration), evaluated with VMC and LRDMC calculations, with the basis sets and constraints defined in Tab.~\ref{tab:basis-dimer}. 
The bonding energy $D_e$ is evaluated as the difference between the energy of the dimer and twice the energy of the monomer. 
In the LRDMC calculations the lattice mesh $a$ is reported in parentheses, with the exception of the $a\to 0$ extrapolation. 
For a comparison, we also report the values obtained from the experiment and from other computational methods. 
The stochastic error for the QMC evaluations is reported only for the bonding energy $D_e$, and it is smaller than $10^{-4}$~a.u. for the energy evaluations, and smaller than $10^{-3}$ for the variance and dipole evaluations (in the reported units). 
\\
}																			
\label{tab:dimer-bonding}
{\footnotesize																			
\begin{tabular}{ | l   c  |  c c c | c c c |  c | }															\hline																			
																			
	&		&	\multicolumn{3}{c|}{\bf monomer }					&	\multicolumn{3}{c|}{\bf dimer}					&	bonding$^b$			\\
method	&	Ref	&	$E$ [H]	&	$VAR(E)$ [H$^2$]	&	$\mu$ [Deb]	&	$E$ [H]	&	$VAR(E)$ [H$^2$]	&	$\mu$ [Deb]	&	$D_e$ [kcal/mol]			\\
\hline
\hline
VMC/JSD/small-basis1	&	this work	&	-17.24637	&	0.295	&	1.886	&	-34.49918	&	0.593	&	2.563	&	4.05	(	4	) \\
VMC/JSD/small-basis2	&	this work	&	-17.24673	&	0.296	&	1.887	&	-34.50027	&	0.592	&	2.559	&	4.27	(	6	) \\
VMC/JSD/small-basis3	&	this work	&	-17.24710	&	0.300	&	1.891	&	-34.50103	&	0.590	&	2.554	&	4.28	(	6	) \\
VMC/JSD/small-basis4	&	this work	&	-17.24738	&	0.294	&	1.888	&	-34.50170	&	0.587	&	2.559	&	4.36	(	7	) \\
VMC/JSD/small-basis5	&	this work	&	-17.24773	&	0.299	&	1.890	&	-34.50264	&	0.588	&	2.555	&	4.50	(	8	) \\
\hline																			
VMC/JSD/large-basis1	&	this work	&	-17.24891	&	0.248	&	1.919	&	-34.50511	&	0.498	&	2.566	&	4.57	(	5	) \\
VMC/JSD/large-basis2	&	this work	&	-17.24894	&	0.247	&	1.855	&	-34.50513	&	0.498	&	2.520	&	4.55	(	5	) \\
VMC/JSD/large-basis3	&	this work	&	-17.24892	&	0.246	&	1.853	&	-34.50541	&	0.496	&	2.520	&	4.75	(	5	) \\
VMC/JSD/large-basis4	&	this work	&	-17.24911	&	0.244	&	1.895	&	-34.50559	&	0.494	&	2.545	&	4.62	(	5	) \\
VMC/JSD/large-basis5	&	this work	&	-17.24908	&	0.244	&	1.882	&	-34.50551	&	0.490	&	2.544	&	4.61	(	5	) \\
\hline																			
VMC/JAGP/large-basis1	&	this work	&	-17.25436	&	0.215	&	1.902	&	-34.51499	&	0.438	&	2.559	&	3.94	(	4	) \\
VMC/JAGP/large-basis2	&	this work	&	-17.25442	&	0.215	&	1.832	&	-34.51524	&	0.438	&	2.511	&	4.01	(	5	) \\
VMC/JAGP/large-basis3	&	this work	&	-17.25442	&	0.214	&	1.835	&	-34.51520	&	0.436	&	2.503	&	4.00	(	10	) \\
VMC/JAGP/large-basis4	&	this work	&	-17.25448	&	0.213	&	1.866	&	-34.51560	&	0.430	&	2.534	&	4.17	(	5	) \\
VMC/JAGP/large-basis5	&	this work	&	-17.25461	&	0.212	&	1.864	&	-34.51586	&	0.432	&	2.541	&	4.17	(	10	) \\
\hline																			
LRDMC(0.5)/JSD/small-basis1	&	this work	&	-17.26626	&		&	1.854	&	-34.54073	&		&	2.528	&	5.15	(	5	) \\
LRDMC(0.4)/JSD/small-basis1	&	this work	&	-17.26488	&		&	1.869	&	-34.53777	&		&	2.546	&	5.03	(	5	) \\
LRDMC(0.3)/JSD/small-basis1	&	this work	&	-17.26389	&		&	1.877	&	-34.53583	&		&	2.554	&	5.06	(	4	) \\
LRDMC(0.2)/JSD/small-basis1	&	this work	&	-17.26323	&		&	1.879	&	-34.53452	&		&	2.554	&	5.05	(	4	) \\
LRDMC/JSD/small-basis1	&	this work	&	-17.26267	&		&		&	-34.53341	&		&		&	5.06	(	5	) \\
\hline																			
LRDMC(0.5)/JSD/large-basis5	&	this work	&	-17.26475	&		&		&	-34.53737	&		&		&	4.94	(	9	) \\
LRDMC(0.4)/JSD/large-basis5	&	this work	&	-17.26396	&		&		&	-34.53589	&		&		&	5.00	(	9	) \\
LRDMC(0.3)/JSD/large-basis5	&	this work	&	-17.26350	&		&		&	-34.53506	&		&		&	5.06	(	8	) \\
LRDMC(0.2)/JSD/large-basis5	&	this work	&	-17.26318	&		&		&	-34.53424	&		&		&	4.94	(	7	) \\
LRDMC/JSD/large-basis5	&	this work	&	-17.26292	&		&		&	-34.53374	&		&		&	4.95	(	10	) \\
\hline																			
LRDMC(0.5)/JAGP/large-basis5	&	this work	&	-17.26683	&		&		&	-34.54113	&		&		&	4.69	(	5	) \\
LRDMC(0.4)/JAGP/large-basis5	&	this work	&	-17.26621	&		&		&	-34.54005	&		&		&	4.78	(	4	) \\
LRDMC(0.3)/JAGP/large-basis5	&	this work	&	-17.26583	&		&		&	-34.53938	&		&		&	4.85	(	5	) \\
LRDMC(0.2)/JAGP/large-basis5	&	this work	&	-17.26550	&		&		&	-34.53877	&		&		&	4.88	(	6	) \\
LRDMC/JAGP/large-basis5	&	this work	&	-17.26530	&		&		&	-34.53839	&		&		&	4.92	(	7	) \\
\hline																			
BLYP/aug-cc-pVTZ(-f)	&	Ref.~\citenum{Xu:2004fi}	&		&		&	1.810	&		&		&		&	4.18			\\
B3LYP/aug-cc-pVTZ(-f)	&	Ref.~\citenum{Xu:2004fi}	&		&		&	1.856	&		&		&		&	4.57			\\
CCSD(T)/IO275$^a$	&	Ref.~\citenum{Klopper:2000iy}	&		&		&		&		&		&		&	5.02			\\
CCSD(T)/CBS	&	Ref.~\citenum{S22-database}	&		&		&		&		&		&		&	5.02			\\
MP2/CBS	&	Ref.~\citenum{S22-database}	&		&		&		&		&		&		&	5.03			\\
\hline																			
Experiment	&	Refs.~\citenum{Clough:1973bh,Odutola:1980em}	&		&		&	1.855	&		&		&		&	5.44$\pm$0.7 $^c$			\\
\hline
%\hline																			
\multicolumn{9}{p{17cm}}{$^a$ IO275: interaction optimized basis set with 275 basis functions for the H2O dimer., see Ref.~\citenum{Klopper:2000iy}. }\\																			
\multicolumn{9}{p{17cm}}{$^b$ $D_e$ is the total bond energy from the bottom of the well.  }		\\																	
\multicolumn{9}{p{17cm}}{$^c$ The quantity actually measured experimentally is the net bond energy from the lowest vibrational level $D_0$, that is $D_0 = 3.59 \pm 0.5 $ in Ref.~\citenum{Odutola:1980em}; $D_e$ was estimated by adding the zero-point energy calculated at the HF/4-21G level.  }		\\								\end{tabular}																			
}																			
\end{table*}

%dimer bonding - Tab.~\ref{tab:dimer-bonding}
% commenti tab. monomer e dimer
In Tab.~\ref{tab:dimer-bonding} we report the evaluations of the energy, the variance and the total dipole, obtained for the water monomer and dimer 
with the different basis sets of Tab.~\ref{tab:basis-dimer}. 
All computations refers to  
the JSD wave function ansatz,
and the Jastrow correlated antisymmetrized geminal power (JAGP) ansatz, described in Ref.~\citenum{Casula:2003p12694}.
We have considered both the variational Monte Carlo scheme and the fixed-node lattice regularized diffusion Monte Carlo scheme\cite{Casula:2005p14138,Casula:2010p14082}.
We evaluate the bonding energy by considering the difference $D_e$ between the energy of the dimer and twice the energy of the monomer, for each wave function ansatz and QMC scheme.
For a comparison, in Tab.~\ref{tab:dimer-bonding} we also report the experimental evaluations and the results provided by other computational approaches.

In Tab.~\ref{tab:dimer-bonding} it is shown that, at a variational level, the largest basis sets decrease both the energy and the variance, both for the monomer and the dimer.
In particular, the largest variational gains are obtained, in absolute terms, by improving the determinantal part of the wave function: the JSD ansatz with ``small-basisX'' has a variance of $\sim 0.30$~a.u. for the monomer and $\sim 0.59$ for the dimer, that is reduced respectively to $\sim 0.25$ and $\sim 0.49$ for the JSD ansatz with ``large-basisX'', and to $\sim 0.21$ and $\sim 0.43$ for the JAGP ansatz with ``large-basisX''.
However, the improved total  energy and  variance, obtained by switching  from the JSD to the JAGP ansatz, do not necessarily implies  an improvement in the H-bond description: for instance for the large-basis5 the JAGP energy is 5.5~mH lower than the JSD energy for the monomer and 10.3~mH for the dimer, but the evaluation of the H-bond is more accurate for the JSD ansatz ($\sim$4.6~mH) rather than for the JAGP ansatz ($\sim$4.2~mH).
It looks  that JAGP is mainly improving the electronic structure of the monomers, but not the H-bond description. Moreover, in the JAGP ansatz the unphysical charge fluctuations introduced by the AGP part are eliminated only by a very large and in principle complete 
 Jastrow term, see Refs.~\citenum{Sorella:2007p12646,Neuscamman:2012hm}. Therefore, 
 in the JAGP case,  the  evaluation of the H-bond is affected also by the incomplete description 
of the Jastrow with a finite basis set. 
% JSD vmc
Since the JSD ansatz is not affected by this kind of problem, it is easier to obtain reliable descriptions of the H-bond of the water also with a wave function with a relatively small number of parameters, as the one that we have chosen for the VMC-based dynamics.
These considerations have led us to choose the JSD ansatz for the dynamics.
% bonding
Within the JSD ansatz and the VMC scheme, the H-bond evaluation ranges from $\sim$4~kcal/mol for the smallest basis set considered (small-basis1) to $\sim$4.7~kcal/mol for the largest basis sets.  
In terms of absolute energy, the JSD/large-basis5 leads to a decrease of 2.7~mH in the monomer energy, and of 6.3~mH for the dimer energy, with respect to the JSD/small-basis1.

\begin{figure*}[htbp]
\begin{center}
\includegraphics[width=0.7\textwidth]{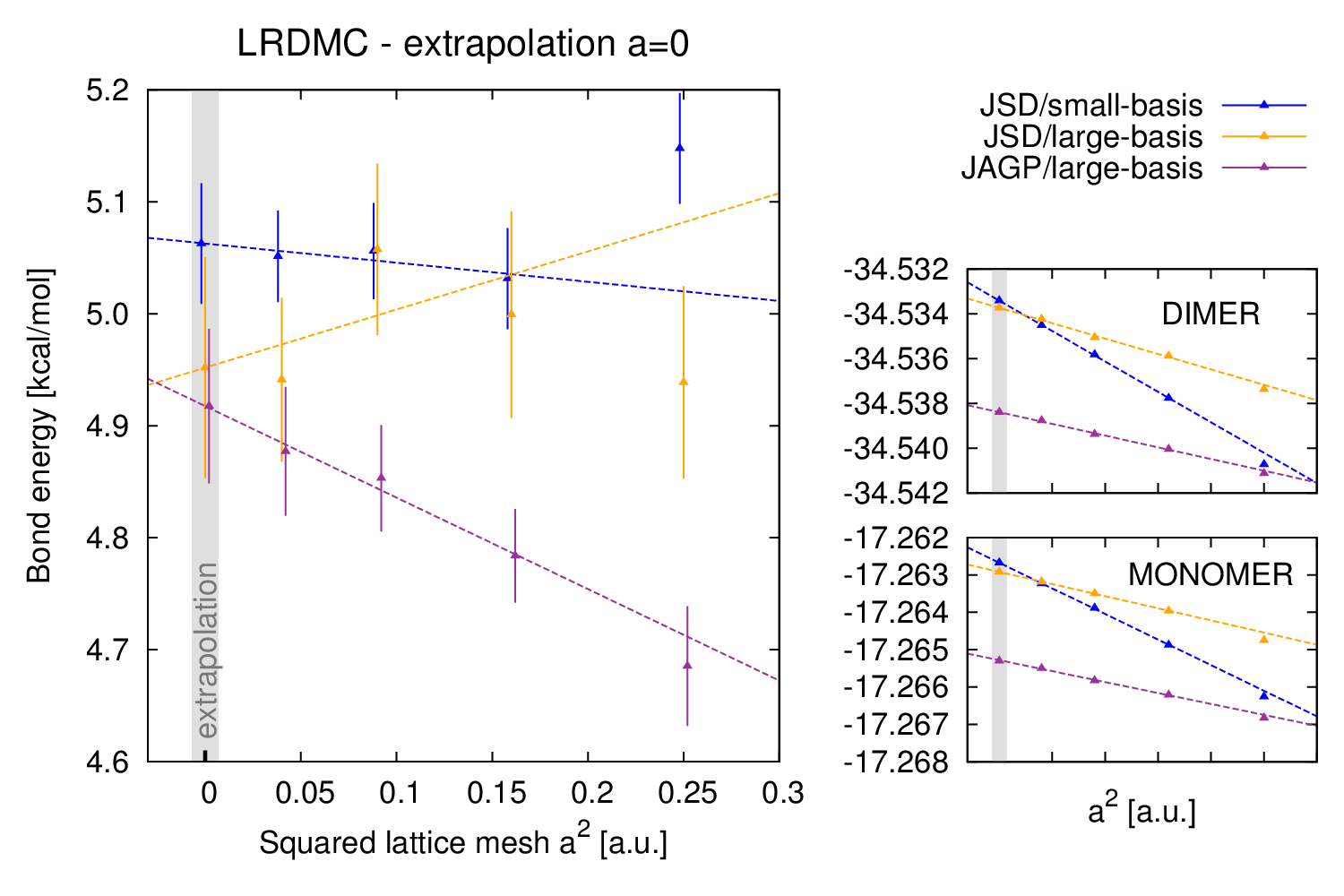}
\caption{
LRDMC extrapolation of the mesh $a\to 0$, for the total energies of the monomer and of the dimer, and of the binding energy. 
The corresponding values are also reported in Tab.~\ref{tab:dimer-bonding}.
The dashed lines correspond to a linear fit of data for the values of the mesh $a$ equal to 0.2, 0.3, 0.4. 
}
\label{fig:lrdmc}
\end{center}
\end{figure*}

% LRDMC
In Tab.~\ref{tab:dimer-bonding} we have also reported the results for the LRDMC calculations of the 
JSD/small-basis1, JSD/large-basis5 and JAGP/large-basis5, 
and
in Fig.~\ref{fig:lrdmc} we show the dependence on the mesh size $a$ of the total energies and of the binding energy.
We observe that the evaluation of $D_e$ is $\sim$5~kcal/mol for all the three wave functions, in agreement with other highly accurate quantum chemical methods such as the CCSD(T) or the MP2, see Refs.~\citenum{Klopper:2000iy,S22-database}.
We also observe that the evaluation of $D_e$ seems not affected by the choice of the LRDMC mesh size $a$ for the JSD ansatz, at least in the range $0.2 \le a \le 0.5$, whereas a small bias with large $a$ can be observed for the JAGP ansatz.
Indeed, in Fig.~\ref{fig:lrdmc} the angular coefficients of the fitting lines for the binding energy is, within the evaluated error of the fitting, compatible with zero for the two JSD calculations, %(respectively -0.2(5) and +0.5(10) for the small and large basis), 
whereas it is not the case for the JAGP wave function. % (-0.8(6)).
The absolute LRDMC energies for the monomer and the dimer, for the extrapolated $a\to 0$, of the JSD/small-basis1 and JSD/large-basis5 ansatzes differ for less than 0.3~mH, indicating that the small-basis1, although much smaller than the large-basis5 and less accurate at the VMC level, provides a nodal surface that seems as good as the one provided by JSD/large-basis5, because 
the  fixed-node projection scheme yields the same electronic correlation for both 
wave functions.

\begin{figure}[htbp]
\begin{center}
\includegraphics[width=0.49\textwidth]{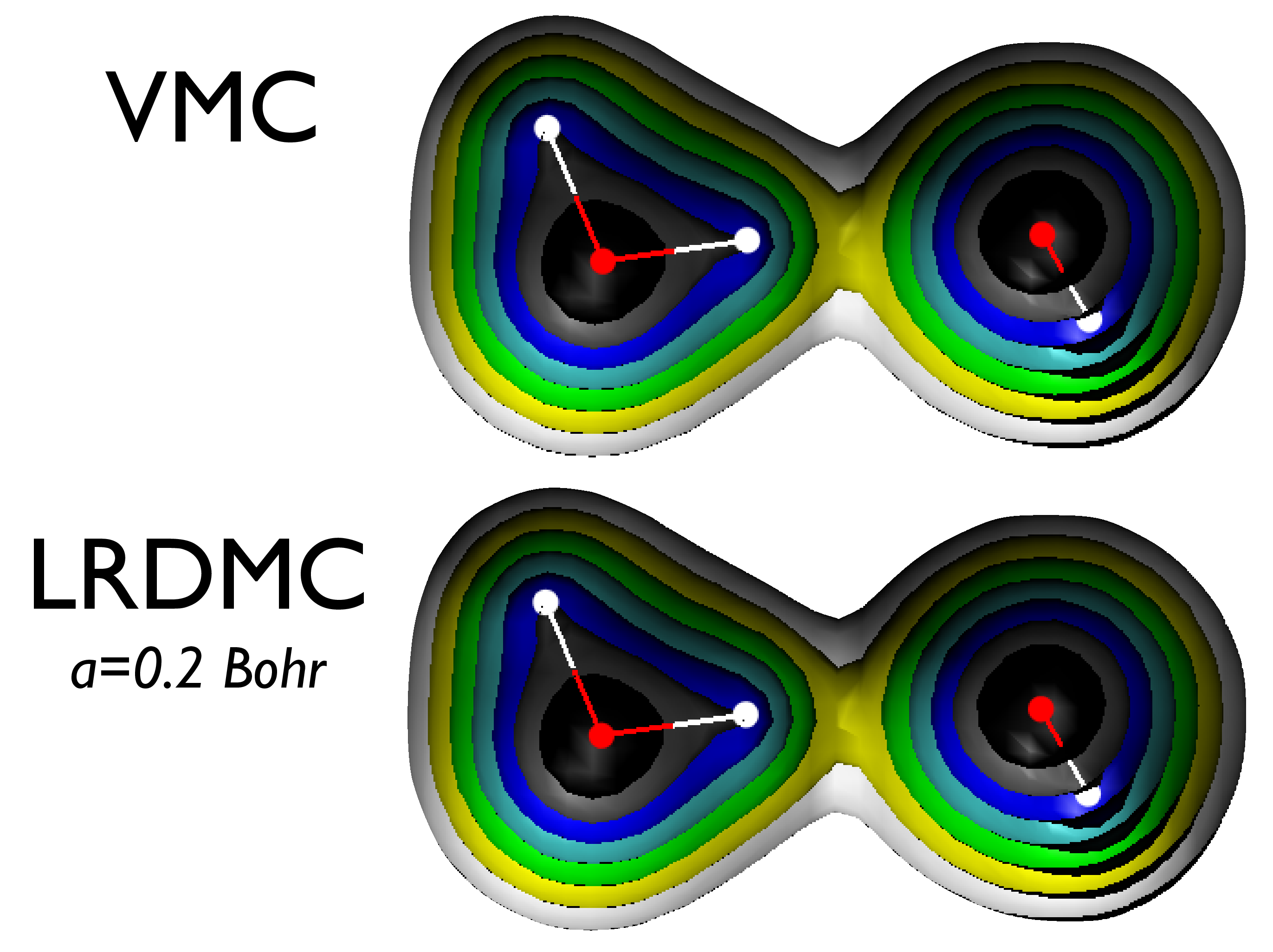}
\caption{
Electronic density of the water dimer, studied with the JSD wave function ansatz and the basis set ``small-basis1'' (see Tab.~\ref{tab:basis-dimer}) which has been used also for the VMC-based molecular dynamics.
The upper picture corresponds to a VMC calculations, and the lower picture to a LRDMC calculation with mesh size $a=0.2$~Bohr.
The reported hypersurfaces are cut in proximity of the plane defined by the donor water molecule, and are colored coded, with white corresponding to a density of 0.01~a.u., yellow to 0.02~a.u., green to 0.04~a.u., cyan to 0.08~a.u., blue to 0.16~a.u., gray to 0.32~a.u., black to 0.64~a.u.
}
\label{fig:density}
\end{center}
\end{figure}

% density
The evaluations of the total dipole $\mu$, for the monomer and the dimer, show that these quantities  are  less affected  than $D_e$ by the wave function ansatz, and we have obtained reliable values for all the considered methods.
In Fig.~\ref{fig:density} we have reported a representation of the electronic density for the dimer, calculated with VMC and with LRDMC(a=0.2).
It can be observed that the electrons distribution is very similar in the two methods.
Moreover, we observe that, in the region between the donor hydrogen $H_d$ and the acceptor oxygen $O_a$, the electronic density is always larger than 0.02~a.u. (yellow hypersurface in Fig.~\ref{fig:density}), due to the presence of the H-bond between the two atoms.

%%%Fig.~\ref{fig:density}: electronic densities for the VMC and LRDMC(a=0.2) calculations, JSD/small-basis1.

%%%%%%%%%%%%%%%%%%%%%%%%%%%%%%%%%%%%%%%%%%%%%%%%%%%%%%%%%%%%%%%%%%%%%%%%%%%%%%%%%%%%%%%%%%%%%
\begin{table*}
\caption{ Geometrical properties (\AA, deg) for the water dimer, see Fig.~\ref{fig:W2dissoc}A, with the VMC/JSD wave function ansatz used for the VMC-based molecular dynamics, in comparison with the results obtained with VMC/JSD and a larger basis, the experimental values, and results obtained from other computational approaches.  }
\label{tab:dimerGEO}
%{\tiny
%{\scriptsize
%{\footnotesize
%{\small
\begin{tabular}{ l c   c c c c c c c  }
%%%
\hline																	
Method	&	Ref.	&	$d(O_aO_d)$	&	$\theta(O_aO_dH_d)$	&	$d(O_dH_d)$	&	$d(O_dH_f)$	&	$d(O_aH_a)$	&	$\theta(H_dO_dH_f)$	&	$\theta(H_aO_aH_a)$	\\
\hline																	
																	
BLYP/aug-cc-pVTZ(-f)	&	Ref.~\citenum{Xu:2004fi}	&	2.952	&	5.9	&	0.981	&	0.971	&	0.973	&	104.8	&	104.7	\\
B3LYP/aug-cc-pVTZ(-f)	&	Ref.~\citenum{Xu:2004fi}	&	2.926	&	5.8	&	0.970	&	0.961	&	0.963	&	105.4	&	105.3	\\
CCSD(T)/IO275$^a$	&	Ref.~\citenum{Klopper:2000iy}	&	2.912	&	5.5	&	0.964	&	0.957	&	0.958	&	104.8	&	104.9	\\
CCSD(T)/cc-pVQZ$^b$	&	Ref.~\citenum{Jurecka:2006bg}	&	2.910	&	4.8	&	0.964	&	0.957	&	0.959	&	104.3	&	104.6	\\
																	
VMC/JSD/small-basis1$^c$	&	this work	&	2.966	&	3.3	&	0.960	&	0.953	&	0.955	&	104.8	&	105.0	\\
VMC/JSD/large-basis5$^d$ 	&	this work	&	2.942	&	4.5	&	0.961	&	0.953	&	0.955	&	105.0	&	105.2	\\

Experiment	&	Ref.~\citenum{Odutola:1980em}	&	2.976 $\pm$ 0.030	&	6$\pm$20	&		&		&		&		&		\\

\hline																	
\multicolumn{9}{l}{$^a$ IO275: interaction optimized basis set with 275 basis functions for the H2O dimer., see Ref.~\citenum{Klopper:2000iy}. }\\																	
\multicolumn{9}{l}{$^b$ Structure used in the S22-database \cite{Jurecka:2006bg,S22-database}. }\\																	
\multicolumn{9}{l}{$^c$ Wave function used for VMC-based MD simulations with 32 and 64 waters in PBC box; see Tab.~\ref{tab:wfpar}. } \\																	
\multicolumn{9}{l}{$^d$ See Tab.~\ref{tab:basis-dimer} }\\																	
%%%
\end{tabular}
%}
\end{table*}
%%%%%%%%%%%%%%%%%%%%%%%%%%%%%%%%%%%%%%%%%%%%%%%%%%%%%%%%%%%%%%%%%%%%%%%%%%%%%%%%%%%%%%%%%%%%%

%Str. opt. - Tab.~\ref{tab:dimerGEO}
The JSD/small-basis1 and JSD/large-basis5 wave functions have also been considered for a structural optimization of the water dimer.
The results are reported in Tab.~\ref{tab:dimerGEO}, and compared with experimental and other ab-initio computational evaluations.
The two wave functions provide structures that are very close, indicating that, at a structural level, the smaller basis do not introduce a large bias, at least at the minimum of the potential energy surface. 
The main difference between them is in the distance between the oxygens, that differs for $\sim$0.02~\AA.
Anyway, both the results are in good agreement with the experimental evaluations and the reported highly accurate quantum chemical calculations.

\begin{figure*}[htbp]
\begin{center}
%{\Large (a)}\includegraphics[width=0.8\textwidth]{basis-wat2/plot_dissociation_abs.eps}\\
%{\Large (b)}\includegraphics[width=0.4\textwidth]{basis-wat2/plot_dissociation.eps}
%{\Large (c)}\includegraphics[width=0.4\textwidth]{basis-wat2/plot_dipole.eps}
{\large A}\includegraphics[width=0.8\textwidth]{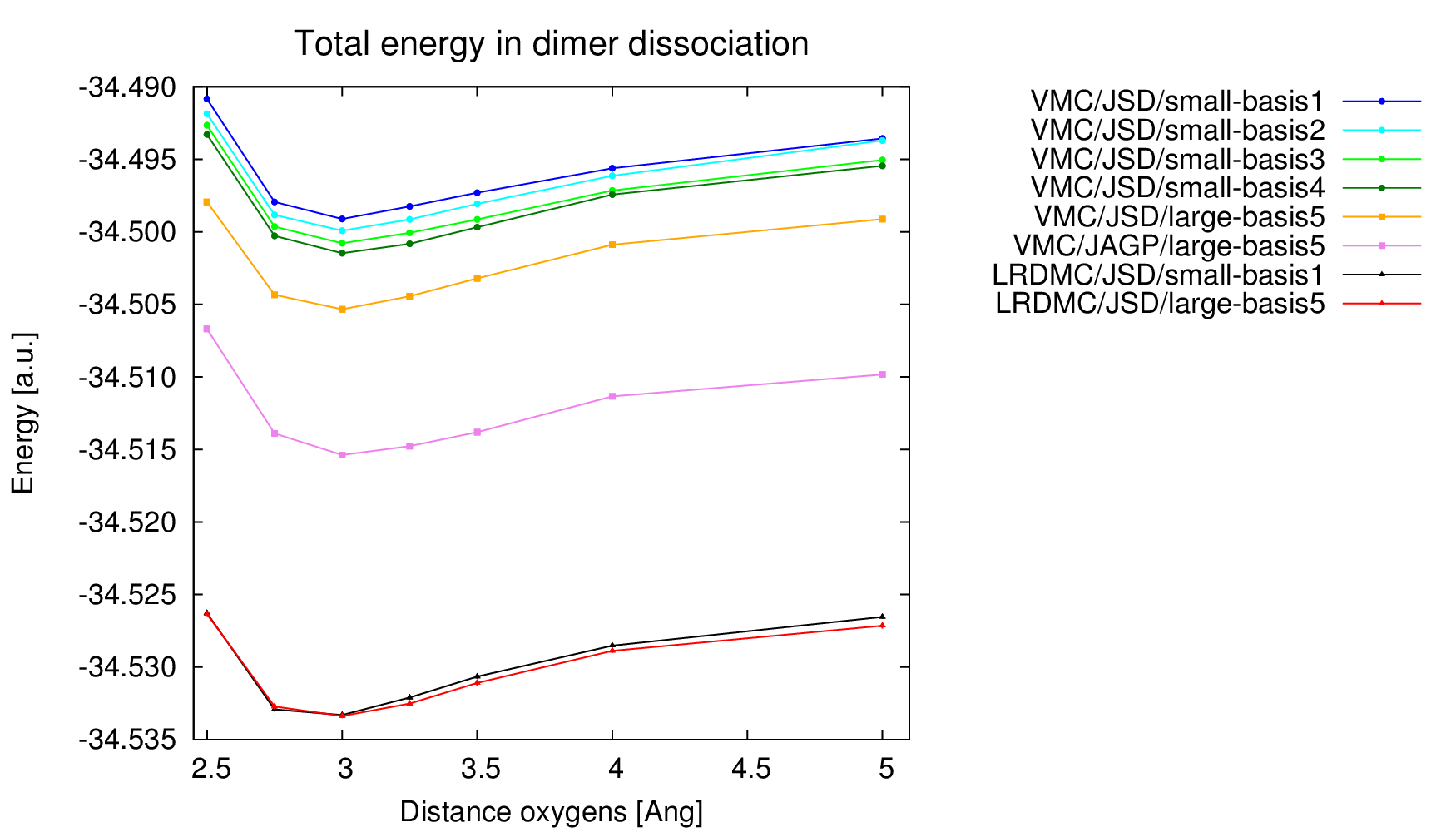}\\
{\large B}\includegraphics[width=0.4\textwidth]{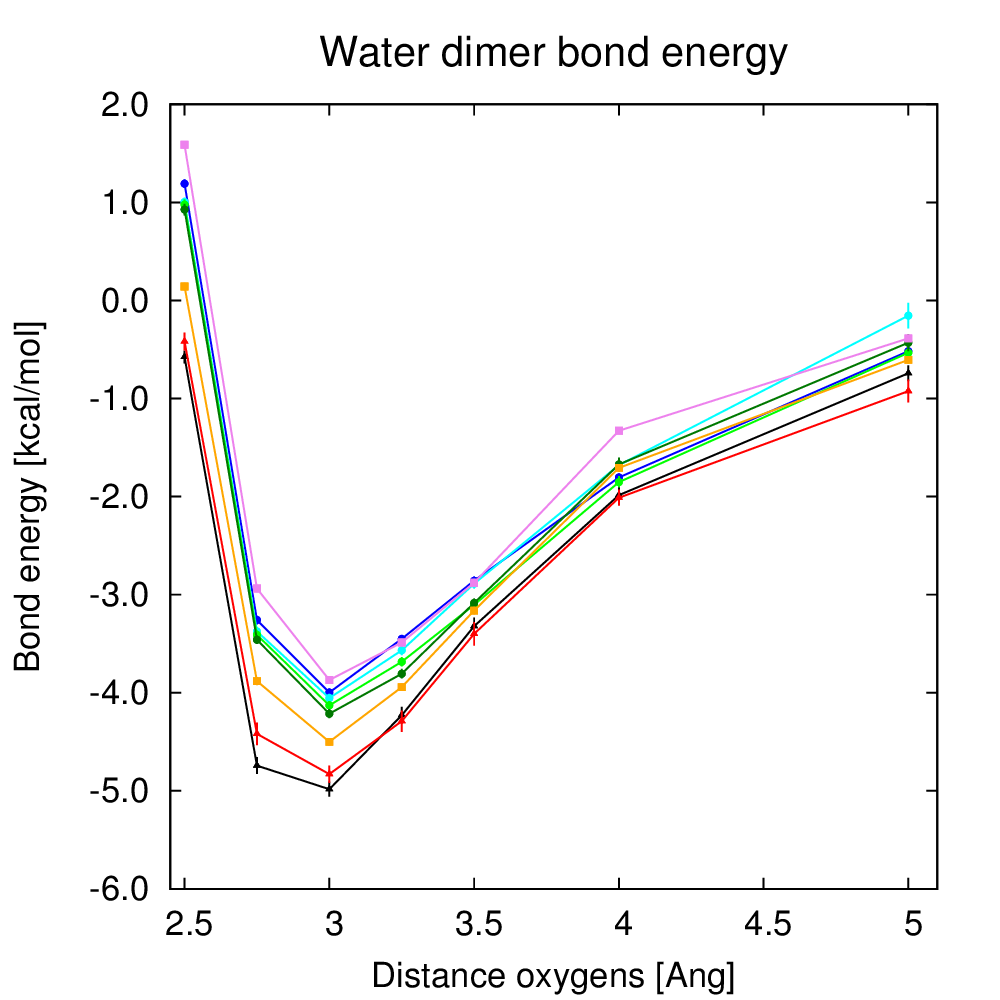}
{\large C}\includegraphics[width=0.4\textwidth]{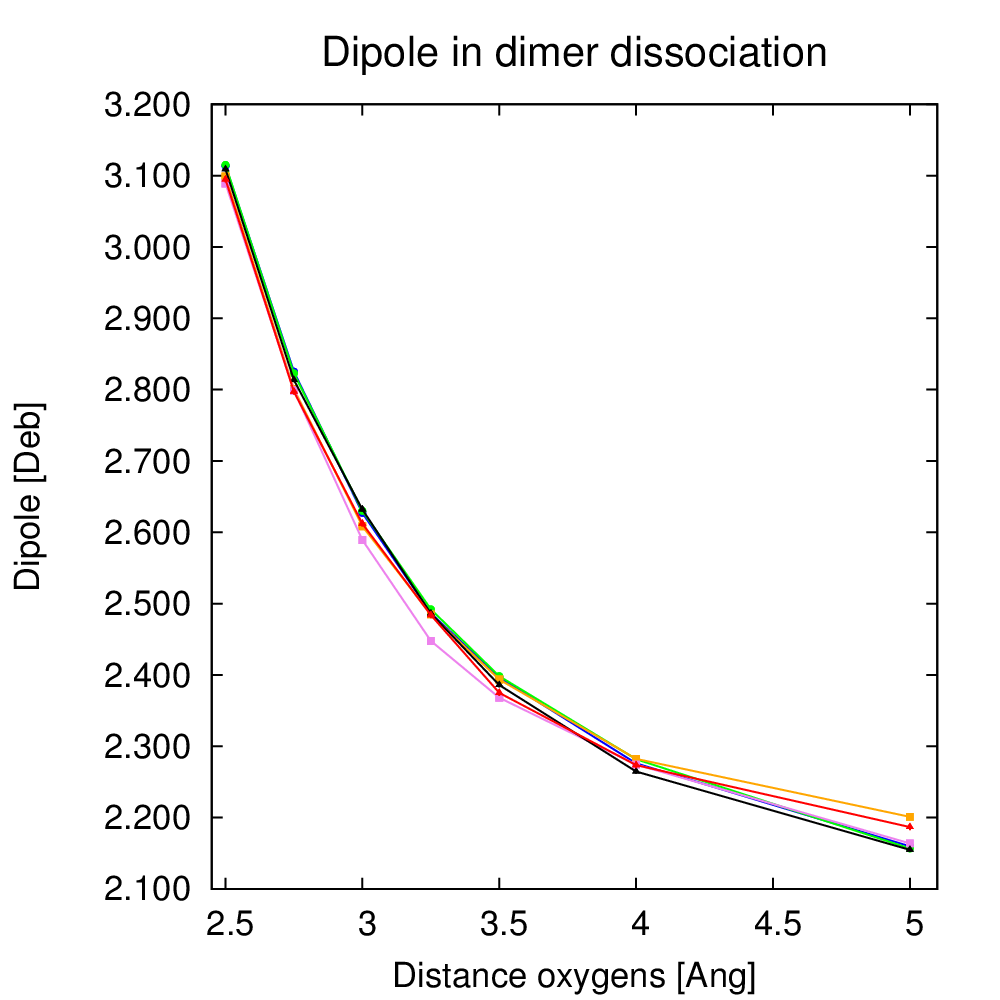}
\caption{
Total energy (panel A), bond energy (panel B) and dipole (panel C) of the water dimer, plotted as a function of the oxygen-oxygen distance $d_{O_dO_a}$, studied with different VMC ansatzes and basis sets, see text and Tab.~\ref{tab:basis-dimer}, using VMC and LRDMC approaches. The LRDMC results are obtained  with a mesh size $a=0.3$~a.u., and the bias due the the mesh size has been corrected by assuming that it is the same in all the configurations, so we have used the corrections obtained from Tab.~\ref{tab:dimer-bonding} 
and Fig.~\ref{fig:lrdmc}.
}
\label{fig:W2dissocApp}
\end{center}
\end{figure*}

%Plot of dissociation:  total energy (a), bond energy (a) and dipole (c) of the water dimer - Fig.~\ref{fig:W2dissocApp}
In order to check our wave function ansatz not only at the structural minimum, but in a larger region of the potential energy surface (PES), we have considered the dissociation of the water dimer. 
In  Fig.~\ref{fig:W2dissocApp} we report the total energy (panel a), bond energy (panel b) and total dipole (panel c) of the water dimer in dissociation.
The dissociation is realized by considering structures with increasing oxygen-oxygen distance, $d_{O_dO_a}$, in the dimer.
At the PES minimum, the donor hydrogen $H_d$ is  rotated of a few degrees from the axis connecting the two oxygens, and slightly moved in the direction of the acceptor oxygen $O_a$ with respect to its equilibrium distance in the monomer, see Tab.~\ref{tab:dimerGEO}. 
However, in order to simplify the dissociation plots reported in Fig.~\ref{fig:W2dissocApp},
 we have considered the two water molecules at exactly their equilibrium configuration, and relatively oriented in order to have $H_d$ in the oxygens axis, as in \citet{Sterpone:2008p12640}.
In Fig.~\ref{fig:W2dissocApp}A we observe that, at the VMC level, each increase in the basis sets  here considered leads to an improved variational energy, and also the JAGP ansatz provides a large variational improvement when compared with the JSD calculation on the same basis.
As expected, the lower energy is obtained at the LRDMC level, for which we have considered both the JSD/small-basis1 and the JSD/large-basis5 ansatz.
The two different trial wave functions appear to yield to the same energies (once the bias for the finite mesh size $a$ has been evaluated and corrected), within the stochastic errors of the evaluations.
In Fig.~\ref{fig:W2dissocApp}B we observe that, at the VMC/JSD level, a larger basis gives essentially a vertical shift of the binding energy, at least for not too large distances. 
This implies that forces should be quite accurate even for the simplest ansatz in the mentioned physical range.
This is an important property with respect to the molecular dynamics, that is sensitive only to the accuracy of forces (slope in the binding)  and not to the absolute value of the binding energy. Indeed, at the variational level,   it appears difficult to define  a Jastrow term  with a reasonably small  number of parameters, that is also able to recover the full  dynamical correlation energy, accessible instead at the LRDMC level, providing the correct  binding of $\sim$5~kcal/mol.
As a further evidence of the quality of our variational approach,  we observe in Fig.~\ref{fig:W2dissocApp}C that the dipole, thus the electronic distribution of the charge, is essentially the same for all the considered methods.

%\clearpage

%%%%%%%%%%%%%%%%%%%%%%%%%%%%%%%%%%%%%%%%%%%%%%%%%%%%%%%%%%%%%%%%%%%%%%%%%%%%%%%%%%%%%%%%%%%%%
\begin{table*}
\caption{ Dissociation energy of water hexamer clusters, calculated as the energy difference between the energy of the water cluster and six times the energy of the monomer. In parenthesis it is reported the energy difference between each cluster and the prism cluster. All the values are in mH. 
The calculations have been done in the geometries optimized with a MP2/aug-cc-pVTZ calculation, taken from Ref.~\citenum{Santra:2008jw}. 
The HF, LDA, PBE, PBE0, BLYP and B3LYP have been obtained using the Orca package, with an aug-cc-pVTZ basis set.
The VMC calculations (in boldface) have been executed with the JSD/small-basis1, as defined in Table~\ref{tab:basis-dimer}, which has been used also for the VMC-based MD simulation. 
The stochastic error of the VMC evaluations are of 0.2~mH for the dissociation energy evaluations, and 0.1~mH for the energy difference with the prism cluster. The highly accurate results obtained using DMC, MP2 and CCSD(T) have been taken from the references reported in the table, to with we refer for computational details.
\\}
\label{tab:hexamers}
%{\tiny
%{\scriptsize
%{\footnotesize
%{\small
\begin{tabular}{ | l c | c | c | c | c | }
\hline																	
Method	&		&	\multicolumn{4}{c|}{ Dissociation Energy [mH] }													\\
	&		&	prism	&	cage			&	book			&	ring			\\
\hline																	
HF	&		&	-41.7	&	-41.9	(	-0.18	)&	-43.7	(	-2.04	)&	-45.7	(	-3.95	)\\
LDA	&		&	-123.3	&	-123.1	(	0.17	)&	-121.7	(	1.63	)&	-117.9	(	5.37	)\\
PBE	&		&	-74.8	&	-75.2	(	-0.40	)&	-76.5	(	-1.70	)&	-76.0	(	-1.21	)\\
PBE0	&		&	-72.0	&	-72.3	(	-0.37	)&	-73.5	(	-1.55	)&	-73.4	(	-1.42	)\\
BLYP	&		&	-60.8	&	-61.6	(	-0.76	)&	-64.0	(	-3.16	)&	-64.6	(	-3.78	)\\
B3LYP	&		&	-65.1	&	-65.7	(	-0.58	)&	-67.6	(	-2.44	)&	-68.0	(	-2.87	)\\
\multicolumn{2}{|c|}{\bf VMC/JSD/small-basis1}		&	{\bf -56.9}	&	{\bf -56.8	(	0.06	)}&	{\bf -56.9	(	-0.09	)}&	{\bf -55.2	(	1.67	)}\\
DMC	&	Ref.~\citenum{Santra:2008jw}	&	-73.2	&	-72.7	(	0.53	)&	-72.3	(	0.90	)&	-70.7	(	2.45	)\\
MP2	&	Ref.~\citenum{Santra:2008jw}	&	-73.3	&	-73.2	(	0.09	)&	-72.8	(	0.46	)&	-71.5	(	1.81	)\\
CCSD(T)	&	Ref.~\citenum{Olson:2007bs}	&	-76.6	&	-76.2	(	0.46	)&	-74.7	(	1.92	)&	-73.3	(	3.33	)\\
CCSD(T)	&	Ref.~\citenum{Gillan:2012jv}	&		&		(	0.39	)&		(	1.12	)&		(	2.70	)\\
\hline																	
\end{tabular}
%}
\end{table*}

\subsection{ Choice of the wave function ansatz }\label{sec.thechoice}

% choice of JSD/small-basis1
Considering that in the VMC-based molecular dynamics we need a stable and compact wave function that can be efficiently, quickly and systematically optimized after every  ion movement, and that in the liquid water every water is surrounded by other four waters, with a distance between the oxygens that may range from $\sim$2.5\AA~ to $\sim$3.5\AA, we have chosen the smallest considered basis, {i.e.} JSD/small-basis1.
We are confident that the vertical energy shift observed in the water dimer with larger basis sets or ansatzes with a larger number of parameters, see Fig.~\ref{fig:W2dissocApp}B, will not affect substantially our results in the liquid water, and that they will be more accurate than the DFT approaches typically used to study the liquid water.
% confronto DFT
Our VMC wave function is indeed a real many-body wave function, which recovers the dynamical electronic correlation with the various terms of the Jastrow factor.  
We have tested it over the water dimer,  in subsection~\ref{sec:dimer-test}, but we think that the improvement of our VMC approach over DFT, in terms of accuracy, is even larger for liquid water, where the packing of the water molecules makes the correlation larger and more challenging. 
%\section{ Water hexamer clusters }
In support of our believe we show, in Tab.~\ref{tab:hexamers}, the dissociation energies of four hexamer clusters of waters, respectively in prism, cage, book and ring configuration,\cite{Santra:2008jw} calculated with the wave function ansatz and the approach that we have used also for our VMC-based molecular dynamics (VMC with JSD/small-basis1), in comparison with a Hartree-Fock calculation, several DFT calculations with commonly used density functionals, and with some highly accurate quantum chemical approaches: namely diffusion Monte Carlo (DMC), second order Moller-Plesset perturbation theory (MP2), and couple cluster with single, double and perturbative triple excitations (CCSD(T)).
In the table we can observe, in agreement with Refs.~\citenum{Santra:2008jw,Gillan:2012jv}, that the typical DFT approaches rank the four hexamers in the wrong way, in relation to their dissociation energy. 
The VMC calculations obtained with the ansatz used also for the dynamics is instead much more reliable of any of the DFT-based calculations, and they are in fair agreement with the most accurate calculations.

% LRDMC
As already observed, the accuracy of  QMC-based calculations are further improved, and indistinguishable from the best known results if the lattice regularized Diffusion Monte Carlo method is applied. 
However, at present the computer time required for the simulation by LRDMC of several molecules --- like the one presented here --- is still out of reach. We have to remark that, when considering geometrical relaxation or dynamics, what is important is not the total energy but the forces between couples of atoms, namely it is crucial to have accurate energy derivatives of the binding energy profile.   As it can be seen in the inset of Fig.~\ref{fig:W2dissoc}, by shifting the binding energy curve in order to have the minimum on the x-axis (the shift not affecting its derivative),  we obtain a rather good description of the binding shape in the relevant region of $R$ between $2.5$\AA~ and $3.2$\AA, and an acceptable error in the large distance region.
On the other hand, experience has shown that geometrical properties, namely the force values around equilibrium distance, are very well determined by the  simple variational ansatz in Eq.(\ref{thewf}), as it is clearly shown in Subsection~\ref{sec:dimer-test}.  The LRDMC usually provides only a substantial correction to the energetics, and therefore it is not expected to play an important role  for static quantities like $g(r)$.

% FIG LRDMC in S22 geometry

\section{Results for Liquid Water} \label{sec.results}

\begin{figure}
\begin{center}
\includegraphics[width=0.99\columnwidth]{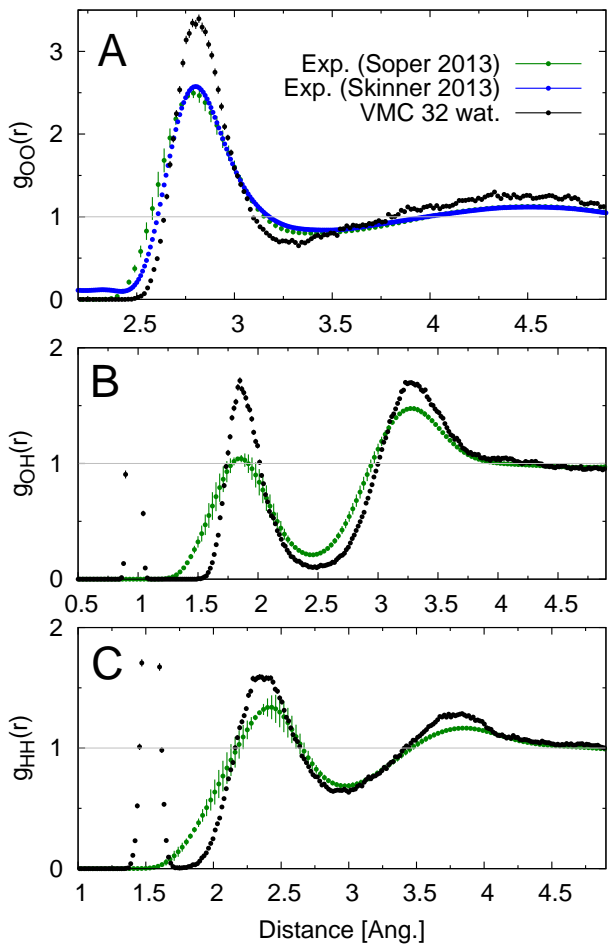}
\caption{
Radial distribution function obtained with 32 waters by a VMC-based dynamics in NVT ensemble (see text) 
as compared with 
the recent X-ray and neutron diffraction experiments of \citet{Skinner:2013cw} and \citet{Soper:2013bs}: %(X-ray\cite{Skinner:2013cw} or neutron diffraction\cite{Soper:2013bs}) 
panel A: Oxygen-Oxygen, panel B: Oxygen-Hydrogen,  panel C:  Hydrogen-Hydrogen. }
A blow-up of the oxygen-oxygen first peak is reported in Fig.~\ref{fig:1stpeak}.
\label{fig:qmc32}
\end{center}
\end{figure}

\begin{figure*}
\begin{center}
\includegraphics[width=0.99\textwidth]{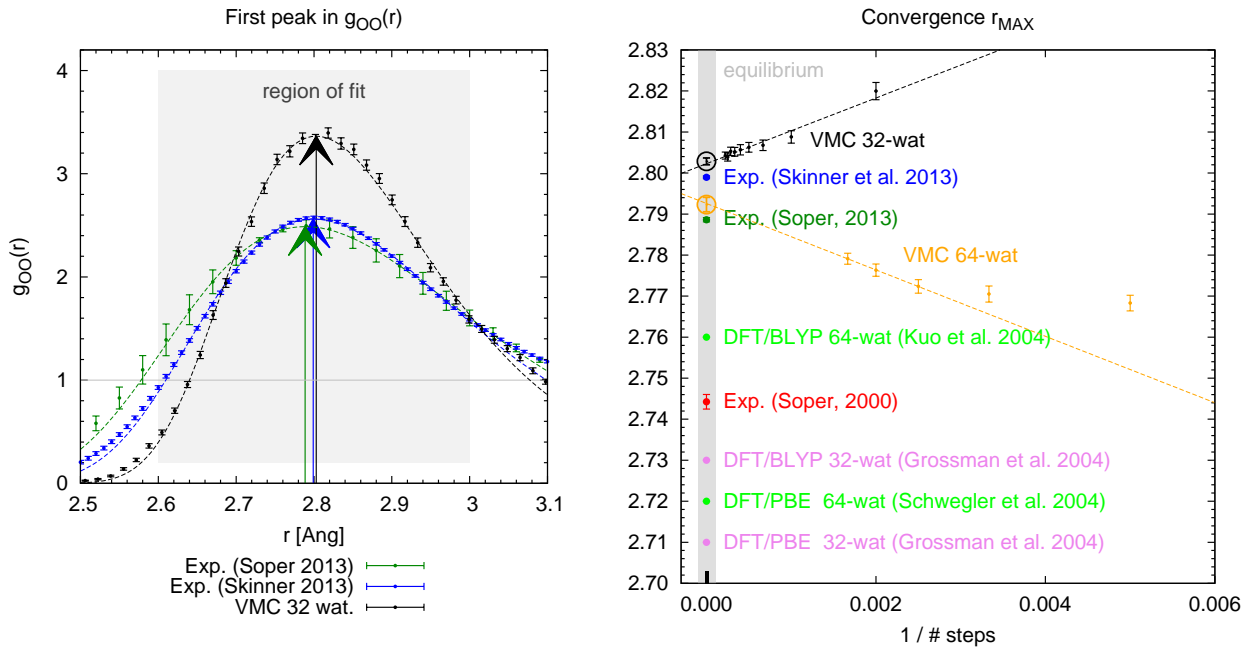}
\caption{ 
Left panel: 
First peak in the oxygen-oxygen radial distribution function 
obtained from X-ray and neutron diffraction experiments\cite{Skinner:2013cw,Soper:2013bs},
%(X-ray\cite{Skinner:2013cw} in blue, neutron diffraction\cite{Soper:2000ik,Soper:2013bs} in green), 
and from from the VMC-based MD simulation with a PBC box of 32 waters in NVT ensemble (in black), having neglected the first 2000 steps.
The $g_{\rm OO}(r)$ data have been fitted with a Gumbel function 
%$f(r) = a \exp( -{r-\mu \over b} -\exp -{ r-\mu \over b})$ 
in the region of r in the range 2.6 -- 3.0 \AA\ (gray square).
The values of $r_\textrm{MAX}$ have been highlighted by arrows with corresponding colors.
Right panel: 
$r_\textrm{MAX}$ as a function of the inverse VMC-MD simulation length  
({\em i.e.} the number of steps). The points outside the gray region refer to time averaged quantities obtained 
without disregarding the initial part of the MD simulation, just to emphasize the smooth approach to equilibrium. Dashed  lines interpolating the leftmost  points are reported. 
The equilibrated values, reported inside the gray region, are estimated by eliminating from the trajectory the first part (namely, considering the trajectory from step 2000 to 4500 for the 32 waters, and from 400 to 600 for the 64 waters). The agreement between these values and the linear extrapolations (dashed lines) shows that at least this quantity is equilibrated 
 within the time simulation length. 
These values are compared with the experimental evaluations and some of the published results\cite{Grossman:2004dw,Kuo:2004p21046,Schwegler:2004ii} for DFT-based approaches (other results from literature are reported in the Tab.~\ref{tab:RDF-OO-review}).  
}
\label{fig:1stpeak}
\end{center}
\end{figure*}

We apply  the molecular dynamics driven by quantum Monte Carlo forces (see Methods), 
introduced recently for the simulation of liquid hydrogen at high pressures\cite{Mazzola:2014dl}.
We have employed a  simulation of 32 waters in the canonical NVT ensemble at ambient temperature $T=300$\,K and experimental density, thus in a cubic cell with box side 
$L=9.86$\,\AA\ % check
and  periodic boundary conditions (PBC). 
Since the values of the atomic masses are not affecting the static equilibrium properties, we have set both the hydrogen and the oxygen masses to 1\,aru,
and we have done about 5000 iterations (that we can estimate to roughly correspond to more than 40 ps of simulation in a standard Newtonian MD simulation), 
where at each iteration we optimize about 
12000 variational parameters with 9 steps of efficient energy optimizations based on 
the so called linear method\cite{cyrus}. We have done several tests\cite{Luo:2014kj} confirming that 
it is possible with this scheme to correctly follow the Born-Oppenheimer energy surface, 
namely the variational wave function 
remains at the minimum possible energy during the time evolution of the  atomic positions. 
The RDFs that we obtain from the VMC-based molecular simulations, having neglected the first 2000 steps of equilibration, are reported in Fig.~\ref{fig:qmc32}, 
in comparison with experimental results.
We have verified that, within this Langevin scheme, the  correlation time estimated by the 
convergence of the RDF %with much longer DFT calculations, 
is less than 2000 iterations, and therefore we 
are confident that our results represent well equilibrated properties.
At variance of the Newtonian dynamics, our advanced method makes use of an appropriate friction 
matrix, which has been proved to be very helpful to reduce the autocorrelation time\cite{Luo:2014kj} as it allows a smooth  approach to the equilibrium (see Fig.~\ref{fig:1stpeak}).

% RDF

% 1st peak

As a starting point of our dynamics we have used equilibrated configurations generated by the DFT molecular dynamics with BLYP functional. 
The BLYP functional describes the water dimer (the simplest system displaying the hydrogen bond) with a reasonable accuracy, comparable with the one  obtained within our VMC scheme, as can be seen in Fig.~\ref{fig:W2dissoc}. 
Nevertheless, the peak positions and shapes of the RDFs are substantially 
different on the target 32 water system.

\begin{figure}[h]
\includegraphics[width=0.45\textwidth]{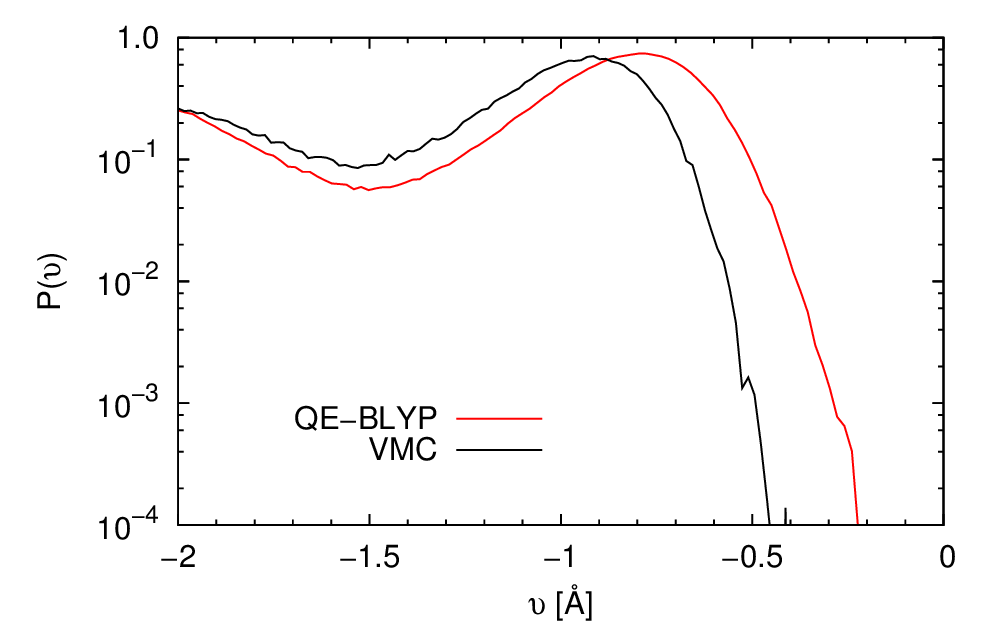}
\caption{Distribution of the proton-transfer coordinate in ab initio dynamics of liquid water at 300~K. 
This coordinate is defined as $\nu = d({\rm O}_{\rm d}\text{-}{\rm H}_{\rm d})-d({\rm O}_{\rm a}\text{-}{\rm H}_{\rm d})$ 
on two water molecules connected by a hydrogen bond (see Fig.~\ref{fig:W2dissoc}A). We haven't observe one event that $\nu > 0$, 
namely no transfer among $\sim 1.8\times 10^{6}$ and $\sim 3.6\times 10^{5}$ hydrogen bonds in BLYP and VMC.}
\label{fig:proton-transfer}
\end{figure}

We see in Fig.~\ref{fig:qmc32} that these first results are very encouraging.
Despite the noise, the outcome is quite clear, because the $g_{\rm OO}(r)$ is much 
closer to experiments than the corresponding DFT calculations.
Not only the radial distribution function is much less overstructured but also, as discussed in the introduction, the position of the first peak is almost indistinguishable from the most recent experiments \cite{Skinner:2013cw,Soper:2013bs}.
At this point it is worth to comment about the error bars in the experimental data.
While the Skinner data are extracted from the x-ray scattering intensities, the Soper data\cite{Soper:2000ik,Soper:PRL2008,Soper:2013bs} are obtained from Empirical Potential Structural Refinement (EPSR) of joint neutron and x-ray data, i.e. they are not bare neutron diffraction data. The large error bars of Soper (2000) are therefore not directly experimental error bars but they refer to a range of different EPSR fits that would model almost equally well the experimental data. Better fits have been published by the same author more recently, see e.g. Ref.~\citenum{Soper:2013bs}, which reports the likely best structural refinement to date.
From the theoretical side the quality of the approximation used for the  electronic correlation  affects the  accuracy  of the RDFs profile. In fact, 
 the first peak's position  has been already improved with respect to standard DFT functionals
 by  employing  the simplest (MP2)  post-Hatree-Fock technique\cite{DelBen:2013ir}.
Moreover quantum effects should broaden the peaks without shifting the corresponding maxima, as it was shown before, within 
DFT, in Ref.~\citenum{Morrone:2008kd}. 
Although this remains, until now, a rather controversial issue\cite{Ceriotti:2013ka,Fritsch:2014hh}, because of the lack of long enough ab-initio simulations with quantum effects included,
our  results seem to support the claim made in Ref.~\cite{Morrone:2008kd} about the relevance of proton quantum corrections in water. 
Indeed  our RDFs for classical ions remain sizably different from experiments, 
as far as the broadening and the heights of the first peaks are concerned,
especially for what concerns the 
$g_{\rm OH}(r)$ and $g_{\rm HH}(r)$ radial distribution functions, where quantum effects are expected to be much more important.
In addition, quantum effects also enhance the the probability of the transient autoprotolysis events, namely proton transfer between water molecules, which were found in a small but nonnegligible fraction by measuring the proton-transfer coordinate\cite{Ceriotti:2013ka}. Consistently, during our classical-ions simulation with VMC, the autoprotolysis event has not been observed, see Fig.~\ref{fig:proton-transfer}.

In order to avoid possible size effects we have studied in Fig.~\ref{fig:1stpeak} the 
position of the first peak with a much shorter simulation ($\sim$600 steps, corresponding to about 
5ps) with 64 molecules.
 Our method  equilibrates rather smoothly with the length of the 
simulation (say, $\# {\rm steps}$), and this nice property, coming from our optimized damped 
MD,  has allowed us to obtain 
a rather well converged value     
of  the peak position also in the 64 water case. 
This further supports the validity of our claim.

\section{Discussion and Conclusion} \label{sec.discussion}

\begin{figure}
\centering
\includegraphics[width=0.9\columnwidth]{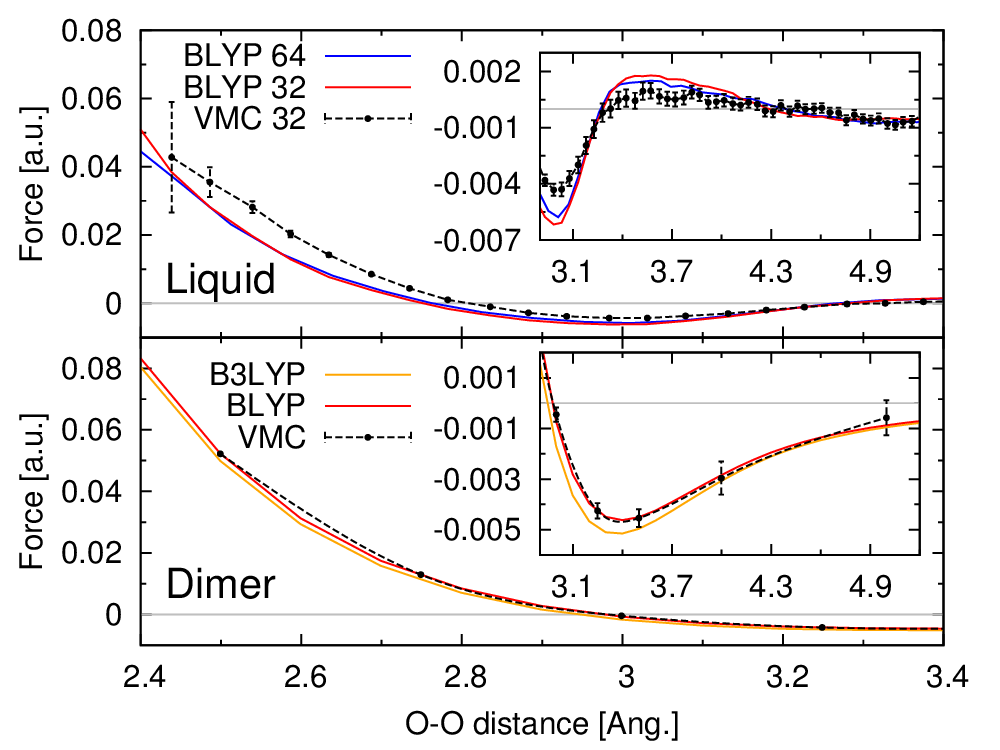}
\caption{The net molecular force difference of a pair of water monomers $(i,j)$ projected on their oxygen-oxygen (O-O) direction 
$({\vec f}^i_{\rm H_2O}-{\vec f}^j_{\rm H_2O})\cdot ({\vec r}_{\rm O_i}-{\vec r}_{\rm O_j})/r_{\rm O_i O_j}$
 as a function of O-O distance $r_{\rm O_i O_j}$ where the net molecular force ${\vec f}_{\rm H_2O}={\vec f}_{\rm H_1}+{\vec f}_{\rm H_2}+{\vec f}_{\rm O}$. 
The upper panel shows that in liquid water the VMC and DFT/BLYP forces have sizable difference at short range below 3\AA\ 
and the difference between two DFT/BLYP calculations with 32 and 64 molecules in the unit cell is negligible. 
The lower panel shows that in the water dimer the two body behavior of VMC, DFT/BLYP and DFT/B3LYP results are almost the same both at short and long distance.}
\label{fig:force_compare}
\end{figure}

In conclusion, we have done the first ab-initio simulation of liquid water by quantum Monte 
Carlo, showing that this technique is now mature to enter the field of ab initio molecular simulations.
This opens a new frontier in water simulations, because several questions about its structure, its electronic properties  and the phase diagram, also difficult to answer experimentally, can  be tackled in the near future thanks to the usage of massive parallel supercomputers and quantum Monte Carlo methods.
We have adopted the most simple quantum Monte Carlo method (the VMC) in a fully consistent and 
controlled way. Despite the roughness of this first attempt (as compared with the most recent 
DFT calculations), a few clear results come out from our study:
\begin{itemize}
% room for improvement
\item The calculation by QMC is feasible albeit computationally heavy,  and there is room for considerable
improvements along this fully ab-initio scheme. For instance it could be possible to work with 
a larger but more accurate basis (see Fig.~\ref{fig:W2dissocApp} and Tab.~\ref{tab:dimer-bonding}) with at most a factor ten  
more computer resources, as our algorithm scales quadratically with the basis dimension.
Moreover, even larger improvements in the QMC-based accuracy are expected when moving from the variational scheme, adopted in this work, to fixed-node projection schemes. Fig.~\ref{fig:lrdmc} shows how both the small and large basis provides a binding energy for the water dimer, evaluated via LRDMC, that are statistically in agreement with highly accurate benchmarks\cite{S22-database,Klopper:2000iy} coming from CCSD(T) (5.02~kcal/mol). Thus, the most convenient choice is given by 
 the smallest basis, allowing a much cheaper wavefunction optimization  during the dynamics, and in principle also an efficient evaluation of the 
 fixed-node diffusion forces for every new nuclear configuration. 
However, the DMC is also affected by  
the finite time step error (in the case of ordinary DMC) or the finite 
lattice mesh (in the case of the LRDMC scheme used here), and the solution of the 
infinite variance problem in DMC (or LRDMC) has  not been clearly solved  yet. 
Therefore we expect that the computational cost for this DMC-based molecular dynamics would be easily two order or magnitude larger than the VMC based method proposed here.  
% better than DFT/GGA
\item The simulated structural properties of liquid water, obtained within our VMC-based molecular dynamics, appears much closer to the experimental observations when compared with  DFT-based molecular dynamics simulations, at least 
within  standard GGA functionals, such as BLYP and PBE.
This is remarkable, because, within our approximation, the two-water interaction was basically 
dealt with the same degree of accuracy of the BLYP functional. As discussed in subsection~\ref{sec.thechoice}, this implies that the accurate description of the many water energy surface, 
and probably the long distance interactions --- usually described within DFT by "ad hoc" strategies --- should be important to close the gap with experiments. 
In this respect, and in order to support our claim on  much simpler systems, we have also verified that our simple variational wave function provides 
a satisfactory description of water hexamers, in good agreement with the most accurate quantum chemical approaches\cite{Santra:2008jw,Gillan:2012jv} --- MP2, DMC, CCSD(T) --- and in contrast with most current DFT functionals 
for  liquid water (see Tab.~\ref{tab:hexamers}).
%
%%%- water hexamer clusters: 
%%%MP2, DMC, CCSD(T): \cite{Santra:2008jw}
%%%see also ref.\cite{Gillan:2012jv,Gillan:2013hq}
%
In order to put further evidence 
about the systematic difference between QMC and DFT with BLYP functional 
we show in Fig.~\ref{fig:force_compare} that the interaction
between two water monomers, namely the repulsive force 
acting on the O-O axis,  is much different from the BLYP prediction when they  are in the 
liquid water  and not in the vacuum.
\item It is clear that our work can help the DFT community to define accurate but also 
consistent functionals, able to reproduce the experimental results, together with a good description of the chemical interactions between water molecules.
\item Our agreement with experiment is rather satisfactory, and could be probably improved 
if the a larger system (64 waters could be sufficient) and  nuclear quantum corrections (not included in this study) would be considered.
Indeed  the height of the first peak $g_{\rm OO}^{\rm MAX}=g_{\rm OO}(r_{\rm MAX})$ is expected to be overestimated by  $\sim 0.3$~\cite{Grossman:2004dw,Kuhne:2009jq} with respect to the converged value in a DFT dynamics with 32 waters.
If we assume that also in QMC we have the same effects, the agreement with
the experimental value should be substantially improved.  
Moreover, the inclusion of nuclear quantum effects appears to reduce further 
the height of the first peak by about $\sim 0.4$ ($\sim 0.24$) 
 if we consider as reference  the  path integral CPMD calculation reported in Ref.~\citenum{Morrone:2008kd} (Ref.~\citenum{Fritsch:2014hh}).
Therefore, in future studies, by taking into account both size effects and the effect of quantum nuclei, it may be possible to have a fully consistent 
ab-initio description of liquid water by QMC.
\end{itemize} 

We finally remark that, thanks to good scaling properties of QMC algorithms with the system size, this work opens promising perspectives for future applications of such high-level ab initio molecular dynamics technique to study the finite temperature properties of complex liquids and materials.

%%% proton transfer

%%% force projection

%\newpage

\begin{acknowledgments}
A.Z and L.G. acknowledge fundings provided by the European Research Council project n. 240624 within the VII Framework Program of the European Union. Computational resources were supplied by CINECA, PRACE infrastructure (Project 2013081585), the Caliban-HPC centre at the University of L'Aquila and the K-computer at AICS Riken.
G.M acknowledges T. K\"{u}hne and J. Vandevondele, and 
A.Z. acknowledges A. Michaelides and R. Potestio
for useful discussions.
\end{acknowledgments}

%\appendix

\clearpage

%\bibliography{../Bibliography}

\begin{thebibliography}{81}
\expandafter\ifx\csname natexlab\endcsname\relax\def\natexlab#1{#1}\fi
\expandafter\ifx\csname bibnamefont\endcsname\relax
  \def\bibnamefont#1{#1}\fi
\expandafter\ifx\csname bibfnamefont\endcsname\relax
  \def\bibfnamefont#1{#1}\fi
\expandafter\ifx\csname citenamefont\endcsname\relax
  \def\citenamefont#1{#1}\fi
\expandafter\ifx\csname url\endcsname\relax
  \def\url#1{\texttt{#1}}\fi
\expandafter\ifx\csname urlprefix\endcsname\relax\def\urlprefix{URL }\fi
\providecommand{\bibinfo}[2]{#2}
\providecommand{\eprint}[2][]{\url{#2}}

\bibitem[{\citenamefont{Car and Parrinello}(1985)}]{carpar}
\bibinfo{author}{\bibfnamefont{R.}~\bibnamefont{Car}} \bibnamefont{and}
  \bibinfo{author}{\bibfnamefont{M.}~\bibnamefont{Parrinello}},
  \bibinfo{journal}{Phys. Rev. Lett.} \textbf{\bibinfo{volume}{55}},
  \bibinfo{pages}{2471} (\bibinfo{year}{1985}).

\bibitem[{\citenamefont{Laasonen et~al.}(1993)\citenamefont{Laasonen, Sprik,
  Parrinello, and Car}}]{LAASONEN:1993p25911}
\bibinfo{author}{\bibfnamefont{K.}~\bibnamefont{Laasonen}},
  \bibinfo{author}{\bibfnamefont{M.}~\bibnamefont{Sprik}},
  \bibinfo{author}{\bibfnamefont{M.}~\bibnamefont{Parrinello}},
  \bibnamefont{and} \bibinfo{author}{\bibfnamefont{R.}~\bibnamefont{Car}},
  \bibinfo{journal}{J. Chem. Phys.} \textbf{\bibinfo{volume}{99}},
  \bibinfo{pages}{9080} (\bibinfo{year}{1993}).

\bibitem[{\citenamefont{Sprik et~al.}(1996)\citenamefont{Sprik, Hutter, and
  Parrinello}}]{Sprik:1996cf}
\bibinfo{author}{\bibfnamefont{M.}~\bibnamefont{Sprik}},
  \bibinfo{author}{\bibfnamefont{J.}~\bibnamefont{Hutter}}, \bibnamefont{and}
  \bibinfo{author}{\bibfnamefont{M.}~\bibnamefont{Parrinello}},
  \bibinfo{journal}{J. Chem. Phys.} \textbf{\bibinfo{volume}{105}},
  \bibinfo{pages}{1142} (\bibinfo{year}{1996}).

\bibitem[{\citenamefont{Ma et~al.}(2012)\citenamefont{Ma, Zhang, and
  Tuckerman}}]{Ma:2012hf}
\bibinfo{author}{\bibfnamefont{Z.}~\bibnamefont{Ma}},
  \bibinfo{author}{\bibfnamefont{Y.}~\bibnamefont{Zhang}}, \bibnamefont{and}
  \bibinfo{author}{\bibfnamefont{M.~E.} \bibnamefont{Tuckerman}},
  \bibinfo{journal}{J. Chem. Phys.} \textbf{\bibinfo{volume}{137}},
  \bibinfo{pages}{044506} (\bibinfo{year}{2012}).

\bibitem[{\citenamefont{K{\"u}hne et~al.}(2009)\citenamefont{K{\"u}hne, Krack,
  and Parrinello}}]{Kuhne:2009jq}
\bibinfo{author}{\bibfnamefont{T.~D.} \bibnamefont{K{\"u}hne}},
  \bibinfo{author}{\bibfnamefont{M.}~\bibnamefont{Krack}}, \bibnamefont{and}
  \bibinfo{author}{\bibfnamefont{M.}~\bibnamefont{Parrinello}},
  \bibinfo{journal}{J. Chem. Theory Comput.} \textbf{\bibinfo{volume}{5}},
  \bibinfo{pages}{235} (\bibinfo{year}{2009}).

\bibitem[{\citenamefont{Sit and Marzari}(2005)}]{Sit:2005cp}
\bibinfo{author}{\bibfnamefont{P.~H.~L.} \bibnamefont{Sit}} \bibnamefont{and}
  \bibinfo{author}{\bibfnamefont{N.}~\bibnamefont{Marzari}},
  \bibinfo{journal}{J. Chem. Phys.} \textbf{\bibinfo{volume}{122}},
  \bibinfo{pages}{204510} (\bibinfo{year}{2005}).

\bibitem[{\citenamefont{Yoo et~al.}(2009)\citenamefont{Yoo, Zeng, and
  Xantheas}}]{Yoo:2009bb}
\bibinfo{author}{\bibfnamefont{S.}~\bibnamefont{Yoo}},
  \bibinfo{author}{\bibfnamefont{X.~C.} \bibnamefont{Zeng}}, \bibnamefont{and}
  \bibinfo{author}{\bibfnamefont{S.~S.} \bibnamefont{Xantheas}},
  \bibinfo{journal}{J. Chem. Phys.} \textbf{\bibinfo{volume}{130}},
  \bibinfo{pages}{221102} (\bibinfo{year}{2009}).

\bibitem[{\citenamefont{VandeVondele et~al.}(2005)\citenamefont{VandeVondele,
  Mohamed, Krack, Hutter, Sprik, and Parrinello}}]{VandeVondele:2005p21042}
\bibinfo{author}{\bibfnamefont{J.}~\bibnamefont{VandeVondele}},
  \bibinfo{author}{\bibfnamefont{F.}~\bibnamefont{Mohamed}},
  \bibinfo{author}{\bibfnamefont{M.}~\bibnamefont{Krack}},
  \bibinfo{author}{\bibfnamefont{J.}~\bibnamefont{Hutter}},
  \bibinfo{author}{\bibfnamefont{M.}~\bibnamefont{Sprik}}, \bibnamefont{and}
  \bibinfo{author}{\bibfnamefont{M.}~\bibnamefont{Parrinello}},
  \bibinfo{journal}{J. Chem. Phys.} \textbf{\bibinfo{volume}{122}},
  \bibinfo{pages}{014515} (\bibinfo{year}{2005}).

\bibitem[{\citenamefont{Lee and Tuckerman}(2006)}]{Lee:2006ge}
\bibinfo{author}{\bibfnamefont{H.-S.} \bibnamefont{Lee}} \bibnamefont{and}
  \bibinfo{author}{\bibfnamefont{M.~E.} \bibnamefont{Tuckerman}},
  \bibinfo{journal}{J. Chem. Phys.} \textbf{\bibinfo{volume}{125}},
  \bibinfo{pages}{154507} (\bibinfo{year}{2006}).

\bibitem[{\citenamefont{Lee and Tuckerman}(2007)}]{Lee:2007hh}
\bibinfo{author}{\bibfnamefont{H.-S.} \bibnamefont{Lee}} \bibnamefont{and}
  \bibinfo{author}{\bibfnamefont{M.~E.} \bibnamefont{Tuckerman}},
  \bibinfo{journal}{J. Chem. Phys.} \textbf{\bibinfo{volume}{126}},
  \bibinfo{pages}{164501} (\bibinfo{year}{2007}).

\bibitem[{\citenamefont{Grossman et~al.}(2004)\citenamefont{Grossman,
  Schwegler, Draeger, Gygi, and Galli}}]{Grossman:2004dw}
\bibinfo{author}{\bibfnamefont{J.~C.} \bibnamefont{Grossman}},
  \bibinfo{author}{\bibfnamefont{E.}~\bibnamefont{Schwegler}},
  \bibinfo{author}{\bibfnamefont{E.~W.} \bibnamefont{Draeger}},
  \bibinfo{author}{\bibfnamefont{F.}~\bibnamefont{Gygi}}, \bibnamefont{and}
  \bibinfo{author}{\bibfnamefont{G.}~\bibnamefont{Galli}}, \bibinfo{journal}{J.
  Chem. Phys.} \textbf{\bibinfo{volume}{120}}, \bibinfo{pages}{300}
  (\bibinfo{year}{2004}).

\bibitem[{\citenamefont{Kuo et~al.}(2004)\citenamefont{Kuo, Mundy, McGrath,
  Siepmann, VandeVondele, Sprik, Hutter, Chen, Klein, Mohamed
  et~al.}}]{Kuo:2004p21046}
\bibinfo{author}{\bibfnamefont{I.}~\bibnamefont{Kuo}},
  \bibinfo{author}{\bibfnamefont{C.}~\bibnamefont{Mundy}},
  \bibinfo{author}{\bibfnamefont{M.}~\bibnamefont{McGrath}},
  \bibinfo{author}{\bibfnamefont{J.}~\bibnamefont{Siepmann}},
  \bibinfo{author}{\bibfnamefont{J.}~\bibnamefont{VandeVondele}},
  \bibinfo{author}{\bibfnamefont{M.}~\bibnamefont{Sprik}},
  \bibinfo{author}{\bibfnamefont{J.}~\bibnamefont{Hutter}},
  \bibinfo{author}{\bibfnamefont{B.}~\bibnamefont{Chen}},
  \bibinfo{author}{\bibfnamefont{M.}~\bibnamefont{Klein}},
  \bibinfo{author}{\bibfnamefont{F.}~\bibnamefont{Mohamed}},
  \bibnamefont{et~al.}, \bibinfo{journal}{J. Phys. Chem. B}
  \textbf{\bibinfo{volume}{108}}, \bibinfo{pages}{12990}
  (\bibinfo{year}{2004}).

\bibitem[{\citenamefont{Todorova et~al.}(2006)\citenamefont{Todorova,
  Seitsonen, Hutter, Kuo, and Mundy}}]{Todorova:2006fr}
\bibinfo{author}{\bibfnamefont{T.}~\bibnamefont{Todorova}},
  \bibinfo{author}{\bibfnamefont{A.~P.} \bibnamefont{Seitsonen}},
  \bibinfo{author}{\bibfnamefont{J.}~\bibnamefont{Hutter}},
  \bibinfo{author}{\bibfnamefont{I.-F.~W.} \bibnamefont{Kuo}},
  \bibnamefont{and} \bibinfo{author}{\bibfnamefont{C.~J.} \bibnamefont{Mundy}},
  \bibinfo{journal}{J. Phys. Chem. B} \textbf{\bibinfo{volume}{110}},
  \bibinfo{pages}{3685} (\bibinfo{year}{2006}).

\bibitem[{\citenamefont{Guidon et~al.}(2008)\citenamefont{Guidon, Schiffmann,
  Hutter, and VandeVondele}}]{Guidon:2008jg}
\bibinfo{author}{\bibfnamefont{M.}~\bibnamefont{Guidon}},
  \bibinfo{author}{\bibfnamefont{F.}~\bibnamefont{Schiffmann}},
  \bibinfo{author}{\bibfnamefont{J.}~\bibnamefont{Hutter}}, \bibnamefont{and}
  \bibinfo{author}{\bibfnamefont{J.}~\bibnamefont{VandeVondele}},
  \bibinfo{journal}{J. Chem. Phys.} \textbf{\bibinfo{volume}{128}},
  \bibinfo{pages}{214104} (\bibinfo{year}{2008}).

\bibitem[{\citenamefont{Zhang et~al.}(2011{\natexlab{a}})\citenamefont{Zhang,
  Donadio, Gygi, and Galli}}]{Zhang:2011hv}
\bibinfo{author}{\bibfnamefont{C.}~\bibnamefont{Zhang}},
  \bibinfo{author}{\bibfnamefont{D.}~\bibnamefont{Donadio}},
  \bibinfo{author}{\bibfnamefont{F.}~\bibnamefont{Gygi}}, \bibnamefont{and}
  \bibinfo{author}{\bibfnamefont{G.}~\bibnamefont{Galli}}, \bibinfo{journal}{J
  Chem. Theory Comput.} \textbf{\bibinfo{volume}{7}}, \bibinfo{pages}{1443}
  (\bibinfo{year}{2011}{\natexlab{a}}).

\bibitem[{\citenamefont{DiStasio et~al.}(2014)\citenamefont{DiStasio, Santra,
  Li, Wu, and Car}}]{DiStasioJr:2014gy}
\bibinfo{author}{\bibfnamefont{R.~A.} \bibnamefont{DiStasio},
  \bibfnamefont{Jr.}},
  \bibinfo{author}{\bibfnamefont{B.}~\bibnamefont{Santra}},
  \bibinfo{author}{\bibfnamefont{Z.}~\bibnamefont{Li}},
  \bibinfo{author}{\bibfnamefont{X.}~\bibnamefont{Wu}}, \bibnamefont{and}
  \bibinfo{author}{\bibfnamefont{R.}~\bibnamefont{Car}}, \bibinfo{journal}{J.
  Chem. Phys.} \textbf{\bibinfo{volume}{141}}, \bibinfo{pages}{084502}
  (\bibinfo{year}{2014}).

\bibitem[{\citenamefont{Grimme}(2004)}]{Grimme:2004db}
\bibinfo{author}{\bibfnamefont{S.}~\bibnamefont{Grimme}}, \bibinfo{journal}{J.
  Comput. Chem.} \textbf{\bibinfo{volume}{25}}, \bibinfo{pages}{1463}
  (\bibinfo{year}{2004}).

\bibitem[{\citenamefont{Grimme}(2006)}]{Grimme:2006fc}
\bibinfo{author}{\bibfnamefont{S.}~\bibnamefont{Grimme}}, \bibinfo{journal}{J.
  Comput. Chem.} \textbf{\bibinfo{volume}{27}}, \bibinfo{pages}{1787}
  (\bibinfo{year}{2006}).

\bibitem[{\citenamefont{von Lilienfeld et~al.}(2004)\citenamefont{von
  Lilienfeld, Tavernelli, Rothlisberger, and
  Sebastiani}}]{vonLilienfeld:2004gz}
\bibinfo{author}{\bibfnamefont{O.}~\bibnamefont{von Lilienfeld}},
  \bibinfo{author}{\bibfnamefont{I.}~\bibnamefont{Tavernelli}},
  \bibinfo{author}{\bibfnamefont{U.}~\bibnamefont{Rothlisberger}},
  \bibnamefont{and}
  \bibinfo{author}{\bibfnamefont{D.}~\bibnamefont{Sebastiani}},
  \bibinfo{journal}{Phys. Rev. Lett.} \textbf{\bibinfo{volume}{93}},
  \bibinfo{pages}{153004} (\bibinfo{year}{2004}).

\bibitem[{\citenamefont{Dion et~al.}(2004)\citenamefont{Dion, Rydberg,
  Schr{\"o}der, Langreth, and Lundqvist}}]{Dion:2004ce}
\bibinfo{author}{\bibfnamefont{M.}~\bibnamefont{Dion}},
  \bibinfo{author}{\bibfnamefont{H.}~\bibnamefont{Rydberg}},
  \bibinfo{author}{\bibfnamefont{E.}~\bibnamefont{Schr{\"o}der}},
  \bibinfo{author}{\bibfnamefont{D.~C.} \bibnamefont{Langreth}},
  \bibnamefont{and} \bibinfo{author}{\bibfnamefont{B.~I.}
  \bibnamefont{Lundqvist}}, \bibinfo{journal}{Phys. Rev. Lett.}
  \textbf{\bibinfo{volume}{92}}, \bibinfo{pages}{246401}
  (\bibinfo{year}{2004}).

\bibitem[{\citenamefont{Schmidt et~al.}(2009)\citenamefont{Schmidt,
  VandeVondele, Kuo, Sebastiani, Siepmann, Hutter, and
  Mundy}}]{Schmidt:2009p20318}
\bibinfo{author}{\bibfnamefont{J.}~\bibnamefont{Schmidt}},
  \bibinfo{author}{\bibfnamefont{J.}~\bibnamefont{VandeVondele}},
  \bibinfo{author}{\bibfnamefont{I.~F.~W.} \bibnamefont{Kuo}},
  \bibinfo{author}{\bibfnamefont{D.}~\bibnamefont{Sebastiani}},
  \bibinfo{author}{\bibfnamefont{J.~I.} \bibnamefont{Siepmann}},
  \bibinfo{author}{\bibfnamefont{J.}~\bibnamefont{Hutter}}, \bibnamefont{and}
  \bibinfo{author}{\bibfnamefont{C.~J.} \bibnamefont{Mundy}},
  \bibinfo{journal}{J. Phys. Chem. B} \textbf{\bibinfo{volume}{113}},
  \bibinfo{pages}{11959} (\bibinfo{year}{2009}).

\bibitem[{\citenamefont{Lin et~al.}(2009)\citenamefont{Lin, Seitsonen,
  Coutinho-Neto, Tavernelli, and Rothlisberger}}]{Lin:2009gm}
\bibinfo{author}{\bibfnamefont{I.-C.} \bibnamefont{Lin}},
  \bibinfo{author}{\bibfnamefont{A.~P.} \bibnamefont{Seitsonen}},
  \bibinfo{author}{\bibfnamefont{M.~D.} \bibnamefont{Coutinho-Neto}},
  \bibinfo{author}{\bibfnamefont{I.}~\bibnamefont{Tavernelli}},
  \bibnamefont{and}
  \bibinfo{author}{\bibfnamefont{U.}~\bibnamefont{Rothlisberger}},
  \bibinfo{journal}{J. Phys. Chem. B} \textbf{\bibinfo{volume}{113}},
  \bibinfo{pages}{1127} (\bibinfo{year}{2009}).

\bibitem[{\citenamefont{Lin et~al.}(2012)\citenamefont{Lin, Seitsonen,
  Tavernelli, and Rothlisberger}}]{Lin:2012es}
\bibinfo{author}{\bibfnamefont{I.-C.} \bibnamefont{Lin}},
  \bibinfo{author}{\bibfnamefont{A.~P.} \bibnamefont{Seitsonen}},
  \bibinfo{author}{\bibfnamefont{I.}~\bibnamefont{Tavernelli}},
  \bibnamefont{and}
  \bibinfo{author}{\bibfnamefont{U.}~\bibnamefont{Rothlisberger}},
  \bibinfo{journal}{J. Chem. Theory Comput.} \textbf{\bibinfo{volume}{8}},
  \bibinfo{pages}{3902} (\bibinfo{year}{2012}).

\bibitem[{\citenamefont{Wang et~al.}(2011)\citenamefont{Wang,
  Rom{\'a}n-P{\'e}rez, Soler, Artacho, and Fern{\'a}ndez-Serra}}]{Wang:2011hx}
\bibinfo{author}{\bibfnamefont{J.}~\bibnamefont{Wang}},
  \bibinfo{author}{\bibfnamefont{G.}~\bibnamefont{Rom{\'a}n-P{\'e}rez}},
  \bibinfo{author}{\bibfnamefont{J.~M.} \bibnamefont{Soler}},
  \bibinfo{author}{\bibfnamefont{E.}~\bibnamefont{Artacho}}, \bibnamefont{and}
  \bibinfo{author}{\bibfnamefont{M.~V.} \bibnamefont{Fern{\'a}ndez-Serra}},
  \bibinfo{journal}{J. Chem. Phys.} \textbf{\bibinfo{volume}{134}},
  \bibinfo{pages}{024516} (\bibinfo{year}{2011}).

\bibitem[{\citenamefont{Zhang et~al.}(2011{\natexlab{b}})\citenamefont{Zhang,
  Wu, Galli, and Gygi}}]{Zhang:2011jj}
\bibinfo{author}{\bibfnamefont{C.}~\bibnamefont{Zhang}},
  \bibinfo{author}{\bibfnamefont{J.}~\bibnamefont{Wu}},
  \bibinfo{author}{\bibfnamefont{G.}~\bibnamefont{Galli}}, \bibnamefont{and}
  \bibinfo{author}{\bibfnamefont{F.}~\bibnamefont{Gygi}}, \bibinfo{journal}{J.
  Chem. Theory Comput.} \textbf{\bibinfo{volume}{7}}, \bibinfo{pages}{3054}
  (\bibinfo{year}{2011}{\natexlab{b}}).

\bibitem[{\citenamefont{Morales et~al.}(2014)\citenamefont{Morales, Gergely,
  McMinis, McMahon, Kim, and Ceperley}}]{Morales:2014jt}
\bibinfo{author}{\bibfnamefont{M.~A.} \bibnamefont{Morales}},
  \bibinfo{author}{\bibfnamefont{J.~R.} \bibnamefont{Gergely}},
  \bibinfo{author}{\bibfnamefont{J.}~\bibnamefont{McMinis}},
  \bibinfo{author}{\bibfnamefont{J.~M.} \bibnamefont{McMahon}},
  \bibinfo{author}{\bibfnamefont{J.}~\bibnamefont{Kim}}, \bibnamefont{and}
  \bibinfo{author}{\bibfnamefont{D.~M.} \bibnamefont{Ceperley}},
  \bibinfo{journal}{J. Chem. Theory Comput.} p.
  \bibinfo{pages}{140512175104006} (\bibinfo{year}{2014}).

\bibitem[{\citenamefont{Morrone and Car}(2008)}]{Morrone:2008kd}
\bibinfo{author}{\bibfnamefont{J.}~\bibnamefont{Morrone}} \bibnamefont{and}
  \bibinfo{author}{\bibfnamefont{R.}~\bibnamefont{Car}},
  \bibinfo{journal}{Phys. Rev. Lett.} \textbf{\bibinfo{volume}{101}},
  \bibinfo{pages}{017801} (\bibinfo{year}{2008}).

\bibitem[{\citenamefont{Habershon et~al.}(2009)\citenamefont{Habershon,
  Markland, and Manolopoulos}}]{Habershon:2009bh}
\bibinfo{author}{\bibfnamefont{S.}~\bibnamefont{Habershon}},
  \bibinfo{author}{\bibfnamefont{T.~E.} \bibnamefont{Markland}},
  \bibnamefont{and} \bibinfo{author}{\bibfnamefont{D.~E.}
  \bibnamefont{Manolopoulos}}, \bibinfo{journal}{J. Chem. Phys.}
  \textbf{\bibinfo{volume}{131}}, \bibinfo{pages}{024501}
  (\bibinfo{year}{2009}).

\bibitem[{\citenamefont{Paesani and Voth}(2009)}]{Paesani:2009ea}
\bibinfo{author}{\bibfnamefont{F.}~\bibnamefont{Paesani}} \bibnamefont{and}
  \bibinfo{author}{\bibfnamefont{G.~A.} \bibnamefont{Voth}},
  \bibinfo{journal}{J. Phys. Chem. B} \textbf{\bibinfo{volume}{113}},
  \bibinfo{pages}{5702} (\bibinfo{year}{2009}).

\bibitem[{\citenamefont{Paesani et~al.}(2010)\citenamefont{Paesani, Yoo,
  Bakker, and Xantheas}}]{Paesani:2010fe}
\bibinfo{author}{\bibfnamefont{F.}~\bibnamefont{Paesani}},
  \bibinfo{author}{\bibfnamefont{S.}~\bibnamefont{Yoo}},
  \bibinfo{author}{\bibfnamefont{H.~J.} \bibnamefont{Bakker}},
  \bibnamefont{and} \bibinfo{author}{\bibfnamefont{S.~S.}
  \bibnamefont{Xantheas}}, \bibinfo{journal}{J. Phys. Chem. Lett.}
  \textbf{\bibinfo{volume}{1}}, \bibinfo{pages}{2316} (\bibinfo{year}{2010}).

\bibitem[{\citenamefont{Li et~al.}(2011)\citenamefont{Li, Walker, and
  Michaelides}}]{Li:2011fd}
\bibinfo{author}{\bibfnamefont{X.~Z.} \bibnamefont{Li}},
  \bibinfo{author}{\bibfnamefont{B.}~\bibnamefont{Walker}}, \bibnamefont{and}
  \bibinfo{author}{\bibfnamefont{A.}~\bibnamefont{Michaelides}},
  \bibinfo{journal}{Proc. Natl. Acad. Sci. U. S. A.}
  \textbf{\bibinfo{volume}{108}}, \bibinfo{pages}{6369} (\bibinfo{year}{2011}).

\bibitem[{\citenamefont{Ceriotti et~al.}(2013)\citenamefont{Ceriotti, Cuny,
  Parrinello, and Manolopoulos}}]{Ceriotti:2013ka}
\bibinfo{author}{\bibfnamefont{M.}~\bibnamefont{Ceriotti}},
  \bibinfo{author}{\bibfnamefont{J.}~\bibnamefont{Cuny}},
  \bibinfo{author}{\bibfnamefont{M.}~\bibnamefont{Parrinello}},
  \bibnamefont{and} \bibinfo{author}{\bibfnamefont{D.~E.}
  \bibnamefont{Manolopoulos}}, \bibinfo{journal}{Proc. Natl. Acad. Sci. U. S.
  A.} \textbf{\bibinfo{volume}{110}}, \bibinfo{pages}{15591}
  (\bibinfo{year}{2013}).

\bibitem[{\citenamefont{Skinner et~al.}(2013)\citenamefont{Skinner, Huang,
  Schlesinger, Pettersson, Nilsson, and Benmore}}]{Skinner:2013cw}
\bibinfo{author}{\bibfnamefont{L.~B.} \bibnamefont{Skinner}},
  \bibinfo{author}{\bibfnamefont{C.}~\bibnamefont{Huang}},
  \bibinfo{author}{\bibfnamefont{D.}~\bibnamefont{Schlesinger}},
  \bibinfo{author}{\bibfnamefont{L.~G.~M.} \bibnamefont{Pettersson}},
  \bibinfo{author}{\bibfnamefont{A.}~\bibnamefont{Nilsson}}, \bibnamefont{and}
  \bibinfo{author}{\bibfnamefont{C.~J.} \bibnamefont{Benmore}},
  \bibinfo{journal}{J. Chem. Phys.} \textbf{\bibinfo{volume}{138}},
  \bibinfo{pages}{074506} (\bibinfo{year}{2013}).

\bibitem[{\citenamefont{Soper}(2013)}]{Soper:2013bs}
\bibinfo{author}{\bibfnamefont{A.~K.} \bibnamefont{Soper}},
  \bibinfo{journal}{ISRN Phys. Chem.} \textbf{\bibinfo{volume}{2013}},
  \bibinfo{pages}{1} (\bibinfo{year}{2013}).

\bibitem[{\citenamefont{Soper}(2000)}]{Soper:2000ik}
\bibinfo{author}{\bibfnamefont{A.~K.} \bibnamefont{Soper}},
  \bibinfo{journal}{Chem. Phys.} \textbf{\bibinfo{volume}{258}},
  \bibinfo{pages}{121} (\bibinfo{year}{2000}).

\bibitem[{\citenamefont{Coccia et~al.}(2014)\citenamefont{Coccia, Varsano, and
  Guidoni}}]{Coccia:2014do}
\bibinfo{author}{\bibfnamefont{E.}~\bibnamefont{Coccia}},
  \bibinfo{author}{\bibfnamefont{D.}~\bibnamefont{Varsano}}, \bibnamefont{and}
  \bibinfo{author}{\bibfnamefont{L.}~\bibnamefont{Guidoni}},
  \bibinfo{journal}{J. Chem. Theory Comput.} p.
  \bibinfo{pages}{140114080315003} (\bibinfo{year}{2014}).

\bibitem[{\citenamefont{Zen et~al.}(2012)\citenamefont{Zen, Zhelyazov, and
  Guidoni}}]{AZenJCTC2012}
\bibinfo{author}{\bibfnamefont{A.}~\bibnamefont{Zen}},
  \bibinfo{author}{\bibfnamefont{D.}~\bibnamefont{Zhelyazov}},
  \bibnamefont{and} \bibinfo{author}{\bibfnamefont{L.}~\bibnamefont{Guidoni}},
  \bibinfo{journal}{J. Chem. Theory Comput.} \textbf{\bibinfo{volume}{8}},
  \bibinfo{pages}{4204} (\bibinfo{year}{2012}).

\bibitem[{\citenamefont{Luo et~al.}(2014)\citenamefont{Luo, Zen, and
  Sorella}}]{Luo:2014kj}
\bibinfo{author}{\bibfnamefont{Y.}~\bibnamefont{Luo}},
  \bibinfo{author}{\bibfnamefont{A.}~\bibnamefont{Zen}}, \bibnamefont{and}
  \bibinfo{author}{\bibfnamefont{S.}~\bibnamefont{Sorella}},
  \bibinfo{journal}{J. Chem. Phys.} \textbf{\bibinfo{volume}{141}},
  \bibinfo{pages}{194112} (\bibinfo{year}{2014}).

\bibitem[{\citenamefont{Mazzola et~al.}(2014)\citenamefont{Mazzola, Yunoki, and
  Sorella}}]{Mazzola:2014dl}
\bibinfo{author}{\bibfnamefont{G.}~\bibnamefont{Mazzola}},
  \bibinfo{author}{\bibfnamefont{S.}~\bibnamefont{Yunoki}}, \bibnamefont{and}
  \bibinfo{author}{\bibfnamefont{S.}~\bibnamefont{Sorella}},
  \bibinfo{journal}{Nat. Commun.} \textbf{\bibinfo{volume}{5}},
  \bibinfo{pages}{3487} (\bibinfo{year}{2014}).

\bibitem[{\citenamefont{Mazzola and Sorella}(2014)}]{Mazzola:2014wc}
\bibinfo{author}{\bibfnamefont{G.}~\bibnamefont{Mazzola}} \bibnamefont{and}
  \bibinfo{author}{\bibfnamefont{S.}~\bibnamefont{Sorella}},
  \bibinfo{journal}{Phys. Rev. Lett.} \textbf{\bibinfo{volume}{114}},
  \bibinfo{pages}{105701} (\bibinfo{year}{2015}).

\bibitem[{\citenamefont{Foulkes et~al.}(2001)\citenamefont{Foulkes, Mitas,
  Needs, and Rajagopal}}]{Foulkes:2001p19717}
\bibinfo{author}{\bibfnamefont{W.~M.~C.} \bibnamefont{Foulkes}},
  \bibinfo{author}{\bibfnamefont{L.}~\bibnamefont{Mitas}},
  \bibinfo{author}{\bibfnamefont{R.~J.} \bibnamefont{Needs}}, \bibnamefont{and}
  \bibinfo{author}{\bibfnamefont{G.}~\bibnamefont{Rajagopal}},
  \bibinfo{journal}{Rev. Mod. Phys.} \textbf{\bibinfo{volume}{73}},
  \bibinfo{pages}{33} (\bibinfo{year}{2001}).

\bibitem[{\citenamefont{Grossman and Mitas}(2005)}]{Grossman:2005p21519}
\bibinfo{author}{\bibfnamefont{J.}~\bibnamefont{Grossman}} \bibnamefont{and}
  \bibinfo{author}{\bibfnamefont{L.}~\bibnamefont{Mitas}},
  \bibinfo{journal}{Phys. Rev. Lett.} \textbf{\bibinfo{volume}{94}},
  \bibinfo{pages}{056403} (\bibinfo{year}{2005}).

\bibitem[{\citenamefont{Attaccalite and Sorella}(2008)}]{attaccalite}
\bibinfo{author}{\bibfnamefont{C.}~\bibnamefont{Attaccalite}} \bibnamefont{and}
  \bibinfo{author}{\bibfnamefont{S.}~\bibnamefont{Sorella}},
  \bibinfo{journal}{Phys. Rev. Lett.} \textbf{\bibinfo{volume}{100}},
  \bibinfo{pages}{114501} (\bibinfo{year}{2008}).

\bibitem[{\citenamefont{Mazzola et~al.}(2012)\citenamefont{Mazzola, Zen, and
  Sorella}}]{mazzola_finite-temperature_2012}
\bibinfo{author}{\bibfnamefont{G.}~\bibnamefont{Mazzola}},
  \bibinfo{author}{\bibfnamefont{A.}~\bibnamefont{Zen}}, \bibnamefont{and}
  \bibinfo{author}{\bibfnamefont{S.}~\bibnamefont{Sorella}},
  \bibinfo{journal}{J. Chem. Phys.} \textbf{\bibinfo{volume}{137}},
  \bibinfo{pages}{134112} (\bibinfo{year}{2012}), ISSN
  \bibinfo{issn}{0021-9606, 1089-7690}.

\bibitem[{\citenamefont{Reboredo and Kim}(2014)}]{reboredo2014generalizing}
\bibinfo{author}{\bibfnamefont{F.~A.} \bibnamefont{Reboredo}} \bibnamefont{and}
  \bibinfo{author}{\bibfnamefont{J.}~\bibnamefont{Kim}}, \bibinfo{journal}{J.
  Chem. Phys.} \textbf{\bibinfo{volume}{140}}, \bibinfo{pages}{074103}
  (\bibinfo{year}{2014}).

\bibitem[{\citenamefont{Ceperley and Dewing}(1999)}]{ceperley_penalty_1999}
\bibinfo{author}{\bibfnamefont{D.~M.} \bibnamefont{Ceperley}} \bibnamefont{and}
  \bibinfo{author}{\bibfnamefont{M.}~\bibnamefont{Dewing}},
  \bibinfo{journal}{The Journal of Chemical Physics}
  \textbf{\bibinfo{volume}{110}}, \bibinfo{pages}{9812} (\bibinfo{year}{1999}),
  ISSN \bibinfo{issn}{0021-9606, 1089-7690}.

\bibitem[{qmc(June 27, 2013)}]{qmcwastrue}
\emph{\bibinfo{title}{See s. sorella's talk at the ``advances in quantum monte
  carlo techniques for non-relativistic many-body systems''}},
  \bibinfo{howpublished}{\url{http://www.int.washington.edu/talks/WorkShops/int_13_2a/}}
  (\bibinfo{year}{June 27, 2013}).

\bibitem[{\citenamefont{K{\"u}hne et~al.}(2007)\citenamefont{K{\"u}hne, Krack,
  Mohamed, and Parrinello}}]{Kuhne:2007df}
\bibinfo{author}{\bibfnamefont{T.}~\bibnamefont{K{\"u}hne}},
  \bibinfo{author}{\bibfnamefont{M.}~\bibnamefont{Krack}},
  \bibinfo{author}{\bibfnamefont{F.}~\bibnamefont{Mohamed}}, \bibnamefont{and}
  \bibinfo{author}{\bibfnamefont{M.}~\bibnamefont{Parrinello}},
  \bibinfo{journal}{Phys. Rev. Lett.} \textbf{\bibinfo{volume}{98}},
  \bibinfo{pages}{066401} (\bibinfo{year}{2007}).

\bibitem[{\citenamefont{Schwegler et~al.}(2004)\citenamefont{Schwegler,
  Grossman, Gygi, and Galli}}]{Schwegler:2004ii}
\bibinfo{author}{\bibfnamefont{E.}~\bibnamefont{Schwegler}},
  \bibinfo{author}{\bibfnamefont{J.~C.} \bibnamefont{Grossman}},
  \bibinfo{author}{\bibfnamefont{F.}~\bibnamefont{Gygi}}, \bibnamefont{and}
  \bibinfo{author}{\bibfnamefont{G.}~\bibnamefont{Galli}}, \bibinfo{journal}{J.
  Chem. Phys.} \textbf{\bibinfo{volume}{121}}, \bibinfo{pages}{5400}
  (\bibinfo{year}{2004}).

\bibitem[{\citenamefont{Drummond et~al.}(2004)\citenamefont{Drummond, Towler,
  and Needs}}]{Drummond:2004p18505}
\bibinfo{author}{\bibfnamefont{N.~D.} \bibnamefont{Drummond}},
  \bibinfo{author}{\bibfnamefont{M.~D.} \bibnamefont{Towler}},
  \bibnamefont{and} \bibinfo{author}{\bibfnamefont{R.~J.} \bibnamefont{Needs}},
  \bibinfo{journal}{Phys. Rev. B} \textbf{\bibinfo{volume}{70}},
  \bibinfo{pages}{235119} (\bibinfo{year}{2004}).

\bibitem[{\citenamefont{Zen et~al.}(2013)\citenamefont{Zen, Luo, Sorella, and
  Guidoni}}]{zenwater}
\bibinfo{author}{\bibfnamefont{A.}~\bibnamefont{Zen}},
  \bibinfo{author}{\bibfnamefont{Y.}~\bibnamefont{Luo}},
  \bibinfo{author}{\bibfnamefont{S.}~\bibnamefont{Sorella}}, \bibnamefont{and}
  \bibinfo{author}{\bibfnamefont{L.}~\bibnamefont{Guidoni}},
  \bibinfo{journal}{J. Chem. Theory Comput.} \textbf{\bibinfo{volume}{9}},
  \bibinfo{pages}{4332} (\bibinfo{year}{2013}).

\bibitem[{\citenamefont{Casula et~al.}(2005)\citenamefont{Casula, Filippi, and
  Sorella}}]{Casula:2005p14138}
\bibinfo{author}{\bibfnamefont{M.}~\bibnamefont{Casula}},
  \bibinfo{author}{\bibfnamefont{C.}~\bibnamefont{Filippi}}, \bibnamefont{and}
  \bibinfo{author}{\bibfnamefont{S.}~\bibnamefont{Sorella}},
  \bibinfo{journal}{Phys. Rev. Lett.} \textbf{\bibinfo{volume}{95}},
  \bibinfo{pages}{100201} (\bibinfo{year}{2005}).

\bibitem[{\citenamefont{Casula et~al.}(2010)\citenamefont{Casula, Moroni,
  Sorella, and Filippi}}]{Casula:2010p14082}
\bibinfo{author}{\bibfnamefont{M.}~\bibnamefont{Casula}},
  \bibinfo{author}{\bibfnamefont{S.}~\bibnamefont{Moroni}},
  \bibinfo{author}{\bibfnamefont{S.}~\bibnamefont{Sorella}}, \bibnamefont{and}
  \bibinfo{author}{\bibfnamefont{C.}~\bibnamefont{Filippi}},
  \bibinfo{journal}{J. Chem. Phys.} \textbf{\bibinfo{volume}{132}},
  \bibinfo{pages}{154113} (\bibinfo{year}{2010}).

\bibitem[{\citenamefont{Marchi et~al.}(2009)\citenamefont{Marchi, Azadi,
  Casula, and Sorella}}]{Marchi:2009p12614}
\bibinfo{author}{\bibfnamefont{M.}~\bibnamefont{Marchi}},
  \bibinfo{author}{\bibfnamefont{S.}~\bibnamefont{Azadi}},
  \bibinfo{author}{\bibfnamefont{M.}~\bibnamefont{Casula}}, \bibnamefont{and}
  \bibinfo{author}{\bibfnamefont{S.}~\bibnamefont{Sorella}},
  \bibinfo{journal}{J. Chem. Phys.} \textbf{\bibinfo{volume}{131}},
  \bibinfo{pages}{154116} (\bibinfo{year}{2009}).

\bibitem[{\citenamefont{Sorella}()}]{TurboRVB}
\bibinfo{author}{\bibfnamefont{S.}~\bibnamefont{Sorella}},
  \emph{\bibinfo{title}{{\em TurboRVB} quantum monte carlo package (accessed
  date may 2013)}},
  \urlprefix\url{http://people.sissa.it/~sorella/web/index.html}.

\bibitem[{\citenamefont{Attaccalite}(2005)}]{AttaccalitePhD}
\bibinfo{author}{\bibfnamefont{C.}~\bibnamefont{Attaccalite}}, Ph.D. thesis,
  \bibinfo{school}{SISSA of Trieste} (\bibinfo{year}{2005}).

\bibitem[{\citenamefont{Sorella et~al.}(2011)\citenamefont{Sorella, Casula,
  Spanu, and Dal~Corso}}]{Sorella:2011p24127}
\bibinfo{author}{\bibfnamefont{S.}~\bibnamefont{Sorella}},
  \bibinfo{author}{\bibfnamefont{M.}~\bibnamefont{Casula}},
  \bibinfo{author}{\bibfnamefont{L.}~\bibnamefont{Spanu}}, \bibnamefont{and}
  \bibinfo{author}{\bibfnamefont{A.}~\bibnamefont{Dal~Corso}},
  \bibinfo{journal}{Phys. Rev. B} \textbf{\bibinfo{volume}{83}},
  \bibinfo{pages}{075119} (\bibinfo{year}{2011}).

\bibitem[{\citenamefont{Petruzielo et~al.}(2011)\citenamefont{Petruzielo,
  Toulouse, and Umrigar}}]{Petruzielo:2011p24345}
\bibinfo{author}{\bibfnamefont{F.~R.} \bibnamefont{Petruzielo}},
  \bibinfo{author}{\bibfnamefont{J.}~\bibnamefont{Toulouse}}, \bibnamefont{and}
  \bibinfo{author}{\bibfnamefont{C.~J.} \bibnamefont{Umrigar}},
  \bibinfo{journal}{J. Chem. Phys.} \textbf{\bibinfo{volume}{134}},
  \bibinfo{pages}{064104} (\bibinfo{year}{2011}).

\bibitem[{\citenamefont{Casula and Sorella}(2003)}]{Casula:2003p12694}
\bibinfo{author}{\bibfnamefont{M.}~\bibnamefont{Casula}} \bibnamefont{and}
  \bibinfo{author}{\bibfnamefont{S.}~\bibnamefont{Sorella}},
  \bibinfo{journal}{J. Chem. Phys.} \textbf{\bibinfo{volume}{119}},
  \bibinfo{pages}{6500} (\bibinfo{year}{2003}).

\bibitem[{\citenamefont{Sorella et~al.}(2007)\citenamefont{Sorella, Casula, and
  Rocca}}]{Sorella:2007p12646}
\bibinfo{author}{\bibfnamefont{S.}~\bibnamefont{Sorella}},
  \bibinfo{author}{\bibfnamefont{M.}~\bibnamefont{Casula}}, \bibnamefont{and}
  \bibinfo{author}{\bibfnamefont{D.}~\bibnamefont{Rocca}}, \bibinfo{journal}{J.
  Chem. Phys.} \textbf{\bibinfo{volume}{127}}, \bibinfo{pages}{014105}
  (\bibinfo{year}{2007}).

\bibitem[{\citenamefont{Burkatzki et~al.}(2007)\citenamefont{Burkatzki,
  Filippi, and Dolg}}]{Burkatzki:2007p25447}
\bibinfo{author}{\bibfnamefont{M.}~\bibnamefont{Burkatzki}},
  \bibinfo{author}{\bibfnamefont{C.}~\bibnamefont{Filippi}}, \bibnamefont{and}
  \bibinfo{author}{\bibfnamefont{M.}~\bibnamefont{Dolg}}, \bibinfo{journal}{J.
  Chem. Phys.} \textbf{\bibinfo{volume}{126}}, \bibinfo{pages}{234105}
  (\bibinfo{year}{2007}).

\bibitem[{\citenamefont{Bra{\"\i}da et~al.}(2011)\citenamefont{Bra{\"\i}da,
  Toulouse, Caffarel, and Umrigar}}]{Braida:2011p27951}
\bibinfo{author}{\bibfnamefont{B.}~\bibnamefont{Bra{\"\i}da}},
  \bibinfo{author}{\bibfnamefont{J.}~\bibnamefont{Toulouse}},
  \bibinfo{author}{\bibfnamefont{M.}~\bibnamefont{Caffarel}}, \bibnamefont{and}
  \bibinfo{author}{\bibfnamefont{C.~J.} \bibnamefont{Umrigar}},
  \bibinfo{journal}{J. Chem. Phys.} \textbf{\bibinfo{volume}{134}},
  \bibinfo{pages}{084108} (\bibinfo{year}{2011}).

\bibitem[{\citenamefont{Anderson and Goddard}(2010)}]{Anderson:2010p25299}
\bibinfo{author}{\bibfnamefont{A.~G.} \bibnamefont{Anderson}} \bibnamefont{and}
  \bibinfo{author}{\bibfnamefont{W.~A.} \bibnamefont{Goddard}},
  \bibinfo{journal}{J. Chem. Phys.} \textbf{\bibinfo{volume}{132}},
  \bibinfo{pages}{164110} (\bibinfo{year}{2010}).

\bibitem[{\citenamefont{Zimmerman et~al.}(2009)\citenamefont{Zimmerman,
  Toulouse, Zhang, Musgrave, and Umrigar}}]{Zimmerman:2009hh}
\bibinfo{author}{\bibfnamefont{P.~M.} \bibnamefont{Zimmerman}},
  \bibinfo{author}{\bibfnamefont{J.}~\bibnamefont{Toulouse}},
  \bibinfo{author}{\bibfnamefont{Z.}~\bibnamefont{Zhang}},
  \bibinfo{author}{\bibfnamefont{C.~B.} \bibnamefont{Musgrave}},
  \bibnamefont{and} \bibinfo{author}{\bibfnamefont{C.~J.}
  \bibnamefont{Umrigar}}, \bibinfo{journal}{J. Chem. Phys.}
  \textbf{\bibinfo{volume}{131}}, \bibinfo{pages}{124103}
  (\bibinfo{year}{2009}).

\bibitem[{\citenamefont{Zen et~al.}(2014)\citenamefont{Zen, Coccia, Luo,
  Sorella, and Guidoni}}]{Zen:2014dh}
\bibinfo{author}{\bibfnamefont{A.}~\bibnamefont{Zen}},
  \bibinfo{author}{\bibfnamefont{E.}~\bibnamefont{Coccia}},
  \bibinfo{author}{\bibfnamefont{Y.}~\bibnamefont{Luo}},
  \bibinfo{author}{\bibfnamefont{S.}~\bibnamefont{Sorella}}, \bibnamefont{and}
  \bibinfo{author}{\bibfnamefont{L.}~\bibnamefont{Guidoni}},
  \bibinfo{journal}{J. Chem. Theory Comput.} \textbf{\bibinfo{volume}{10}},
  \bibinfo{pages}{1048} (\bibinfo{year}{2014}).

\bibitem[{\citenamefont{Zen et~al.}(2015)\citenamefont{Zen, Coccia, Gozem,
  Olivucci, and Guidoni}}]{Zen:2015eu}
\bibinfo{author}{\bibfnamefont{A.}~\bibnamefont{Zen}},
  \bibinfo{author}{\bibfnamefont{E.}~\bibnamefont{Coccia}},
  \bibinfo{author}{\bibfnamefont{S.}~\bibnamefont{Gozem}},
  \bibinfo{author}{\bibfnamefont{M.}~\bibnamefont{Olivucci}}, \bibnamefont{and}
  \bibinfo{author}{\bibfnamefont{L.}~\bibnamefont{Guidoni}},
  \bibinfo{journal}{J. Chem. Theory Comput.} \textbf{\bibinfo{volume}{0}},
  \bibinfo{pages}{null} (\bibinfo{year}{2015}),
  \urlprefix\url{http://dx.doi.org/10.1021/ct501122z}.

\bibitem[{\citenamefont{Jurecka et~al.}(2006)\citenamefont{Jurecka, Sponer,
  Cerny, and Hobza}}]{Jurecka:2006bg}
\bibinfo{author}{\bibfnamefont{P.}~\bibnamefont{Jurecka}},
  \bibinfo{author}{\bibfnamefont{J.}~\bibnamefont{Sponer}},
  \bibinfo{author}{\bibfnamefont{J.}~\bibnamefont{Cerny}}, \bibnamefont{and}
  \bibinfo{author}{\bibfnamefont{P.}~\bibnamefont{Hobza}},
  \bibinfo{journal}{Phys. Chem. Chem. Phys.} \textbf{\bibinfo{volume}{8}},
  \bibinfo{pages}{1985} (\bibinfo{year}{2006}).

\bibitem[{S22()}]{S22-database}
\emph{\bibinfo{title}{S22-benchmark noncovalent complexes}},
  \urlprefix\url{http://www.begdb.com/index.php?action=oneDataset&id=4&state=show&order=ASC&by=name_m&method=}.

\bibitem[{\citenamefont{Xu and Goddard}(2004)}]{Xu:2004fi}
\bibinfo{author}{\bibfnamefont{X.}~\bibnamefont{Xu}} \bibnamefont{and}
  \bibinfo{author}{\bibfnamefont{W.~A.} \bibnamefont{Goddard}},
  \bibinfo{journal}{J. Phys. Chem. A} \textbf{\bibinfo{volume}{108}},
  \bibinfo{pages}{2305} (\bibinfo{year}{2004}).

\bibitem[{\citenamefont{Klopper et~al.}(2000)\citenamefont{Klopper, M~van
  Duijneveldt-van~de Rijdt, and van Duijneveldt}}]{Klopper:2000iy}
\bibinfo{author}{\bibfnamefont{W.}~\bibnamefont{Klopper}},
  \bibinfo{author}{\bibfnamefont{J.~G.~C.} \bibnamefont{M~van
  Duijneveldt-van~de Rijdt}}, \bibnamefont{and}
  \bibinfo{author}{\bibfnamefont{F.~B.} \bibnamefont{van Duijneveldt}},
  \bibinfo{journal}{Phys. Chem. Chem. Phys.} \textbf{\bibinfo{volume}{2}},
  \bibinfo{pages}{2227} (\bibinfo{year}{2000}).

\bibitem[{\citenamefont{Clough et~al.}(1973)\citenamefont{Clough, Beers, Klein,
  and Rothman}}]{Clough:1973bh}
\bibinfo{author}{\bibfnamefont{S.~A.} \bibnamefont{Clough}},
  \bibinfo{author}{\bibfnamefont{Y.}~\bibnamefont{Beers}},
  \bibinfo{author}{\bibfnamefont{G.~P.} \bibnamefont{Klein}}, \bibnamefont{and}
  \bibinfo{author}{\bibfnamefont{L.~S.} \bibnamefont{Rothman}},
  \bibinfo{journal}{J. Chem. Phys.} \textbf{\bibinfo{volume}{59}},
  \bibinfo{pages}{2254} (\bibinfo{year}{1973}).

\bibitem[{\citenamefont{Odutola and Dyke}(1980)}]{Odutola:1980em}
\bibinfo{author}{\bibfnamefont{J.~A.} \bibnamefont{Odutola}} \bibnamefont{and}
  \bibinfo{author}{\bibfnamefont{T.~R.} \bibnamefont{Dyke}},
  \bibinfo{journal}{J. Chem. Phys.} \textbf{\bibinfo{volume}{72}},
  \bibinfo{pages}{5062} (\bibinfo{year}{1980}).

\bibitem[{\citenamefont{Neuscamman}(2012)}]{Neuscamman:2012hm}
\bibinfo{author}{\bibfnamefont{E.}~\bibnamefont{Neuscamman}},
  \bibinfo{journal}{Phys. Rev. Lett.} \textbf{\bibinfo{volume}{109}},
  \bibinfo{pages}{203001} (\bibinfo{year}{2012}).

\bibitem[{\citenamefont{Sterpone et~al.}(2008)\citenamefont{Sterpone, Spanu,
  Ferraro, Sorella, and Guidoni}}]{Sterpone:2008p12640}
\bibinfo{author}{\bibfnamefont{F.}~\bibnamefont{Sterpone}},
  \bibinfo{author}{\bibfnamefont{L.}~\bibnamefont{Spanu}},
  \bibinfo{author}{\bibfnamefont{L.}~\bibnamefont{Ferraro}},
  \bibinfo{author}{\bibfnamefont{S.}~\bibnamefont{Sorella}}, \bibnamefont{and}
  \bibinfo{author}{\bibfnamefont{L.}~\bibnamefont{Guidoni}},
  \bibinfo{journal}{J. Chem. Theory Comput.} \textbf{\bibinfo{volume}{4}},
  \bibinfo{pages}{1428} (\bibinfo{year}{2008}).

\bibitem[{\citenamefont{Santra et~al.}(2008)\citenamefont{Santra, Michaelides,
  Fuchs, Tkatchenko, Filippi, and Scheffler}}]{Santra:2008jw}
\bibinfo{author}{\bibfnamefont{B.}~\bibnamefont{Santra}},
  \bibinfo{author}{\bibfnamefont{A.}~\bibnamefont{Michaelides}},
  \bibinfo{author}{\bibfnamefont{M.}~\bibnamefont{Fuchs}},
  \bibinfo{author}{\bibfnamefont{A.}~\bibnamefont{Tkatchenko}},
  \bibinfo{author}{\bibfnamefont{C.}~\bibnamefont{Filippi}}, \bibnamefont{and}
  \bibinfo{author}{\bibfnamefont{M.}~\bibnamefont{Scheffler}},
  \bibinfo{journal}{J. Chem. Phys.} \textbf{\bibinfo{volume}{129}},
  \bibinfo{pages}{194111} (\bibinfo{year}{2008}).

\bibitem[{\citenamefont{Olson et~al.}(2007)\citenamefont{Olson, Bentz, Kendall,
  Schmidt, and Gordon}}]{Olson:2007bs}
\bibinfo{author}{\bibfnamefont{R.~M.} \bibnamefont{Olson}},
  \bibinfo{author}{\bibfnamefont{J.~L.} \bibnamefont{Bentz}},
  \bibinfo{author}{\bibfnamefont{R.~A.} \bibnamefont{Kendall}},
  \bibinfo{author}{\bibfnamefont{M.~W.} \bibnamefont{Schmidt}},
  \bibnamefont{and} \bibinfo{author}{\bibfnamefont{M.~S.}
  \bibnamefont{Gordon}}, \bibinfo{journal}{J. Chem. Theory Comput.}
  \textbf{\bibinfo{volume}{3}}, \bibinfo{pages}{1312} (\bibinfo{year}{2007}).

\bibitem[{\citenamefont{Gillan et~al.}(2012)\citenamefont{Gillan, Manby,
  Towler, and Alfe}}]{Gillan:2012jv}
\bibinfo{author}{\bibfnamefont{M.~J.} \bibnamefont{Gillan}},
  \bibinfo{author}{\bibfnamefont{F.~R.} \bibnamefont{Manby}},
  \bibinfo{author}{\bibfnamefont{M.~D.} \bibnamefont{Towler}},
  \bibnamefont{and} \bibinfo{author}{\bibfnamefont{D.}~\bibnamefont{Alfe}},
  \bibinfo{journal}{J. Chem. Phys.} \textbf{\bibinfo{volume}{136}},
  \bibinfo{pages}{244105} (\bibinfo{year}{2012}).

\bibitem[{\citenamefont{Umrigar et~al.}(2007)\citenamefont{Umrigar, Toulouse,
  Filippi, Sorella, and Rhenning}}]{cyrus}
\bibinfo{author}{\bibfnamefont{C.~J.} \bibnamefont{Umrigar}},
  \bibinfo{author}{\bibfnamefont{J.}~\bibnamefont{Toulouse}},
  \bibinfo{author}{\bibfnamefont{C.}~\bibnamefont{Filippi}},
  \bibinfo{author}{\bibfnamefont{S.}~\bibnamefont{Sorella}}, \bibnamefont{and}
  \bibinfo{author}{\bibfnamefont{H.}~\bibnamefont{Rhenning}},
  \bibinfo{journal}{Phys. Rev. Lett} \textbf{\bibinfo{volume}{98}},
  \bibinfo{pages}{110201} (\bibinfo{year}{2007}).

\bibitem[{\citenamefont{Soper and C.J.}(2008)}]{Soper:PRL2008}
\bibinfo{author}{\bibfnamefont{A.}~\bibnamefont{Soper}} \bibnamefont{and}
  \bibinfo{author}{\bibfnamefont{B.}~\bibnamefont{C.J.}},
  \bibinfo{journal}{Phys. Rev. Lett.} \textbf{\bibinfo{volume}{101}},
  \bibinfo{pages}{065502} (\bibinfo{year}{2008}).

\bibitem[{\citenamefont{Del~Ben et~al.}(2013)\citenamefont{Del~Ben,
  Sch{\"o}nherr, Hutter, and VandeVondele}}]{DelBen:2013ir}
\bibinfo{author}{\bibfnamefont{M.}~\bibnamefont{Del~Ben}},
  \bibinfo{author}{\bibfnamefont{M.}~\bibnamefont{Sch{\"o}nherr}},
  \bibinfo{author}{\bibfnamefont{J.}~\bibnamefont{Hutter}}, \bibnamefont{and}
  \bibinfo{author}{\bibfnamefont{J.}~\bibnamefont{VandeVondele}},
  \bibinfo{journal}{J. Phys. Chem. Lett.} \textbf{\bibinfo{volume}{4}},
  \bibinfo{pages}{3753} (\bibinfo{year}{2013}).

\bibitem[{\citenamefont{Fritsch et~al.}(2014)\citenamefont{Fritsch, Potestio,
  Donadio, and Kremer}}]{Fritsch:2014hh}
\bibinfo{author}{\bibfnamefont{S.}~\bibnamefont{Fritsch}},
  \bibinfo{author}{\bibfnamefont{R.}~\bibnamefont{Potestio}},
  \bibinfo{author}{\bibfnamefont{D.}~\bibnamefont{Donadio}}, \bibnamefont{and}
  \bibinfo{author}{\bibfnamefont{K.}~\bibnamefont{Kremer}},
  \bibinfo{journal}{J. Chem. Theory Comput.} p.
  \bibinfo{pages}{140123151236004} (\bibinfo{year}{2014}).

\end{thebibliography}

\end{document}